\documentclass[11pt,fleqn,a4paper]{article}
\usepackage{amssymb,latexsym,amsmath,amsfonts}
\usepackage{graphicx}

    \advance\textheight by \topskip

    \numberwithin{equation}{section}

    \def\tr{{\rm tr \,}}
    \def\Re{{\rm Re \,}}
    \def\Im{{\rm Im \,}}

    \newtheorem{theorem}{Theorem}[section]
    \newtheorem{lemma}[theorem]{Lemma}

    \newtheorem{Definition}[theorem]{Definition}
    
    \newtheorem{Remark}[theorem]{Remark}
    \newenvironment{remark}{\begin{Remark}\rm}{\end{Remark}}
    \newtheorem{Example}[theorem]{Example}

    \newenvironment{proof}%
    {\rm \trivlist \item[\hskip \labelsep{\bf Proof. }]}%
    {\hspace*{\fill}$\Box$\endtrivlist}
    \newenvironment{varproof}%
    {\rm \trivlist \item[\hskip \labelsep{\bf Proof}]}%
    {\hspace*{\fill}$\Box$\endtrivlist}

\begin{document}

    \begin{center} \Large\bf
        Universal behavior for averages of characteristic polynomials at the origin of the spectrum
    \end{center}

    \

    \begin{center}
        \large
        M. Vanlessen\footnote{Research Assistant of the Fund for
            Scientific Research -- Flanders (Belgium). Supported by FWO research
            project G.0176.02.}\\
            \normalsize \em
            Department of Mathematics, Katholieke Universiteit Leuven, \\
            Celestijnenlaan 200 B, 3001 Leuven, Belgium \\
            \rm maarten.vanlessen@wis.kuleuven.ac.be
    \end{center}\ \\[1ex]

    \begin{abstract}
    It has been shown by Strahov and Fyodorov that averages of products and ratios of
    characteristic polynomials corresponding to Hermitian matrices
    of a unitary ensemble, involve kernels related to
    orthogonal polynomials and their Cauchy transforms. We will
    show that, for the unitary ensemble $\frac{1}{\hat Z_n}\,|\det
    M|^{2\alpha}e^{-nV(M)}dM$ of $n\times n$ Hermitian matrices,
    these kernels have universal behavior at the
    origin of the spectrum, as $n\to\infty$, in terms of Bessel
    functions. Our approach is based on the characterization of orthogonal
    polynomials together with their Cauchy transforms via a matrix Riemann-Hilbert problem, due to
    Fokas, Its and Kitaev, and on an application of the Deift/Zhou steepest
    descent method for matrix Riemann-Hilbert problems to obtain the asymptotic behavior of the
    Riemann-Hilbert problem.
    \end{abstract}

\section{Introduction}

Characteristic polynomials of random matrices are useful to make
predictions about moments of the Riemann-Zeta function, see
\cite{BrezinHikami1,HughesKeatingOconnel1,HughesKeatingOconnel2,KeatingSnaith}.
Another domain where they are of great value is quantum
chromodynamics, see for example
\cite{AkemannDamgaard1,AkemannDamgaard2,Damgaard,VerbaarschotZahed}.
In this paper we consider characteristic polynomials $\det(x-M)$
of random matrices taken from the following unitary ensemble of
$n\times n$ Hermitian matrices $M$, cf.\ \cite{ADMN1,ADMN2,KV2}
\begin{equation}\label{unitary-ensemble}
    \frac{1}{\hat Z_n}|\det M|^{2\alpha} e^{-n\tr V(M)} dM,
        \qquad \alpha>-1/2.
\end{equation}
Here $dM$ is the associated flat Lebesgue measure on the space of
$n \times n$ Hermitian matrices, and $\hat Z_n$ is a normalization
constant. The confining potential $V$ in (\ref{unitary-ensemble})
is a real valued function with enough increase at infinity, for
example a polynomial of even degree with positive leading
coefficient. This unitary ensemble induces a probability density
function on the $n$ eigenvalues $x_1,\ldots ,x_n$ of $M$, see
\cite{Mehta}
\[
    P^{(n)}(x_1,\ldots
    ,x_n)=\frac{1}{Z_n}\prod_{j=1}^{n}w_n(x_j)\Delta^2(x_1,\ldots ,x_n),
\]
where $\Delta(x_1,\ldots ,x_n)=\prod_{i<j}(x_j-x_i)$ stands for
the Vandermonde determinant, where $Z_n$ is a normalization
constant (the partition function), and where $w_n$ is the
following varying weight on the real line
\begin{equation}\label{definition-weight}
    w_n(x)=|x|^{2\alpha}e^{-nV(x)}.
\end{equation}

The unitary ensemble (\ref{unitary-ensemble}) is relevant in
three-dimensional quantum chromodynamics \cite{VerbaarschotZahed},
and has been investigated before in
\cite{ADMN1,ADMN2,KanFrei,KV2,Nishigaki}, where universal behavior
for local eigenvalue correlations is established in various
regimes of the spectrum, as $n\to\infty$.

\medskip

It is known that averages of products and ratios of characteristic
polynomials are intimately related to orthogonal polynomials and
their Cauchy transforms, see
\cite{BaikDeiftStrahov,BrezinHikami1,FyodorovStrahov,MehtaNormand,StrahovFyodorov}.
Let $\pi_{j,n}(x)=x^j+\cdots$ be the $j$-th degree monic
orthogonal polynomial with respect to $w_n$. There is an integral
representation for the monic orthogonal polynomials, which appears
already in the work of Heine in 1878, see for example
\cite{Szego},
\[
    \pi_{n,n}(x)=\int\ldots\int \prod_{j=1}^n(x-x_j)P^{(n)}(x_1,\ldots
    ,x_n)dx_1\ldots dx_n.
\]
So, the monic orthogonal polynomial $\pi_{n,n}$ can be understood
as the average of the characteristic polynomial $\det(x-M)$ over
the unitary ensemble (\ref{unitary-ensemble})
\[
    \langle \det(x-M) \rangle_M = \pi_{n,n}(x).
\]
Here, the brackets are used to denote the average over the
ensemble (\ref{unitary-ensemble}) of random matrices $M$. A first
generalization of this formula was obtained by Br\'ezin and Hikami
\cite{BrezinHikami1}, and also by Mehta and Normand
\cite{MehtaNormand}. They have derived a determinantal formula for
the average of products of characteristic polynomials in terms of
orthogonal polynomials. A further generalization was obtained by
Fyodorov and Strahov \cite{FyodorovStrahov}, who derived a
determinantal formula for the average of both products and ratios
of characteristic polynomials in terms of both orthogonal
polynomials and their Cauchy transforms. Here, the ratios gave
rise to the Cauchy transforms. For explicit formulas and
streamlined proofs of these results we refer to
\cite{BaikDeiftStrahov}.

Recently, Strahov and Fyodorov \cite{StrahovFyodorov} showed, see
also \cite{BaikDeiftStrahov} for an alternative proof, that the
averages of characteristic polynomials of $n\times n$ Hermitian
matrices, are governed by kernels related to orthogonal
polynomials and their Cauchy transforms
\begin{equation}\label{definition-Cauchy-transform}
    h_{j,n}(z)=\frac{1}{2\pi i}\int
    \frac{\pi_{j,n}(x)}{x-z}w_n(x)dx,\qquad\mbox{for $\Im z\neq
    0$.}
\end{equation}
Namely, kernels $W_{I,n+m}$ made of orthogonal polynomials,
kernels $W_{II,n+m}$ made of both orthogonal polynomials and their
Cauchy transforms, and kernels $W_{III,n+m}$ made of Cauchy
transforms of orthogonal polynomials. See Table
\ref{table:finite-kernels} for the explicit expressions of these
kernels. This connection between the averages of characteristic
polynomials and the three kernels is given by, see
\cite{BaikDeiftStrahov,StrahovFyodorov}
\begin{eqnarray*}
    \lefteqn{
    \left\langle\prod_{i=1}^k
    \det(x_i-M)\det(y_i-M)\right\rangle_M= \frac{(c_{n+k-1,n})^k}{\prod_{j=n}^{n+k-1}c_{j,n}}
    \frac{1}{\Delta(\hat x)\Delta(\hat y)}
    \det\left(W_{I,n+k}(x_i,y_j)\right)_{1\leq i,j\leq k},} \\[3ex]
    \lefteqn{
    \left\langle\prod_{i=1}^k
    \frac{\det(y_i-M)}{\det(x_i-M)}\right\rangle_M=
    (-1)^\frac{k(k-1)}{2}(c_{n-1,n})^k\frac{\Delta(\hat x,\hat y)}{\Delta^2(\hat x)\Delta^2(\hat y)}
    \det\left(W_{II,n}(x_i,y_j)\right)_{1\leq i,j\leq k},}
\end{eqnarray*}
and
\begin{eqnarray*}
    \lefteqn{\left\langle\prod_{i=1}^{2k}
    \frac{1}{\det(x_i-M)}\right\rangle_M}\\[2ex]
    && \quad =\,
        (-1)^k
        \frac{(c_{n-k-1,n})^{k}}{(2k)!}\prod_{l=n-k}^{n-1}c_l
    \sum_{\sigma\in S_{2k}}
    \frac{\det\left(W_{III,n-k}(x_{\sigma(i)},x_{\sigma(k+j)})\right)_{1\leq i,j\leq k}}
    {\Delta(x_{\sigma(1)},\ldots ,x_{\sigma(k)})\Delta(x_{\sigma(k+1)},\ldots
    ,x_{\sigma(2k)})},
\end{eqnarray*}
where $\hat x=(x_1,\ldots ,x_k),\, \hat y=(y_1,\ldots ,y_k)$,
where $c_{j,n}=-2\pi i\gamma_{j,n}^2$ with $\gamma_{j,n}$ the
leading coefficient of the $j$-th degree orthonormal polynomial
with respect to $w_n$, and where $S_{2k}$ is the permutation group
of the index set $\{1,\ldots ,2k\}$. There are also explicit
formulas for averages containing non-equal number of
characteristic polynomials in the numerator and the denominator,
in terms of these kernels, see \cite{StrahovFyodorov} for details.
Strahov and Fyodorov \cite{StrahovFyodorov} used this connection,
together with the Riemann-Hilbert (RH) approach, to establish
universal behavior, as $n\to\infty$, for the averages of
characteristic polynomials of random matrices taken from the
unitary ensemble
\begin{equation}\label{unitary-ensemble-deift}
    \frac{1}{\tilde Z_n}e^{-n\tr V(M)}dM,
\end{equation}
in the bulk of the spectrum.

\begin{table}
\begin{center}
    \begin{tabular}{|l|c|}
        \hline
        \multicolumn{2}{|c|}{} \\[-0,5ex]
        \multicolumn{2}{|c|}{Finite kernels} \\
        \multicolumn{2}{|c|}{} \\[-0,5ex]
        \hline
        & \\[-0,5ex]
        $W_{I,n+m}(\zeta,\eta)$ &
            $\frac{\pi_{n+m,n}(\zeta)\pi_{n+m-1,n}(\eta)-\pi_{n+m-1,n}(\zeta)\pi_{n+m,n}(\eta)}
            {\zeta-\eta}$ \\
        & \\[-0,5ex]
        \hline
        & \\[-0,5ex]
        $W_{II,n+m}(\zeta,\eta)$ &
            $\frac{h_{n+m,n}(\zeta)\pi_{n+m-1,n}(\eta)-h_{n+m-1,n}(\zeta)\pi_{n+m,n}(\eta)}
            {\zeta-\eta}$ \\
        & \\[-0,5ex]
        \hline
        & \\[-0,5ex]
        $W_{III,n+m}(\zeta,\eta)$ &
            $\frac{h_{n+m,n}(\zeta)h_{n+m-1,n}(\eta)-h_{n+m-1,n}(\zeta)h_{n+m,n}(\eta)}
            {\zeta-\eta}$ \\[-0,5ex]
        & \\
        \hline
    \end{tabular}
    \caption{Expressions for the finite kernels $W_{I,n+m}, W_{II,n+m}$ and $W_{III,n+m}$,
        cf. \cite{StrahovFyodorov}.}
    \label{table:finite-kernels}
\end{center}
\end{table}

\medskip

It is the goal of this paper to establish universal behavior as
$n\to\infty$, for the kernels $W_{I,n+m},W_{II,n+m}$ and
$W_{III,n+m}$ (and thus also for the averages of characteristic
polynomials) associated to the unitary ensemble
(\ref{unitary-ensemble}), appropriate scaled at the origin such
that the asymptotic eigenvalue density at the origin is $1$. This
scaling limit is called the {\em origin of the spectrum} by
various authors, see for example
\cite{ADMN1,AkemannFyodorov,KanFrei,KV2}. It will turn out that
this universal behavior is described in terms of the Bessel
kernels given in Table \ref{table:limiting-Bessel-kernels}. For
the case $\alpha=0$, our results agree with those of Strahov and
Fyodorov \cite{StrahovFyodorov}.

The issue of universality at the origin of the spectrum for the
averages of characteristic polynomials, corresponding to Hermitian
matrices of the unitary ensemble (\ref{unitary-ensemble}), was
also considered by Akemann and Fyodorov \cite{AkemannFyodorov}.
They showed, on a physical level of rigor using Shohat's method,
that the asymptotic behavior near the origin, as $n\to\infty$, of
the orthogonal polynomials and their Cauchy transforms are
expressed in terms of Bessel and Hankel functions, see
\cite{AkemannFyodorov} for details. However, explicit expressions
for the universal behavior of the three kernels $W_{I,n+m},
W_{II,n+m}$ and $W_{III,n+m}$ at the origin of the spectrum have
not been given yet, which we will determine on a mathematical
level of rigor using the RH approach, as in
\cite{StrahovFyodorov}.

In \cite{AkemannFyodorov} was assumed that the potential $V$ is an
even polynomial with positive leading coefficient, and that the
spectrum support is only one interval. In this paper, we can allow
$V$ to be quite arbitrary, and assume the following
\begin{eqnarray}
    \label{condition-on-V-eq1}
    \lefteqn{
        \mbox{$V: \mathbb R \to \mathbb R$ is real analytic,}
        } \\[1ex]
    \label{condition-on-V-eq2}
    \lefteqn{
        \lim_{|x|\to\infty}\frac{V(x)}{\log(x^2+1)}=+\infty,
        } \\[1ex]
    \label{condition-on-V-eq3}
    \lefteqn{\psi(0) > 0,}
\end{eqnarray}
where $\psi$ is the density of  the equilibrium measure $\mu_V$ in
the presence of the external field $V$, see
\cite{DKM,DKMVZ2,SaffTotik}. The equilibrium measure $\mu_V$ has
compact support, it is supported on a finite union of intervals
(since $V$ is real analytic), and it is absolutely continuous with
respect to the Lebesgue measure, i.e.\ $d\mu_V(x)=\psi(x)dx$. The
importance of $\mu_V$ lies in the fact that its density $\psi$ is
the limiting mean eigenvalue density of the unitary ensemble
(\ref{unitary-ensemble}). Assumption (\ref{condition-on-V-eq3})
then states that the mean eigenvalue density does not vanish at
the origin.

\begin{table}
\begin{center}
    \begin{tabular}{|l|c|c|}
        \hline
        \multicolumn{2}{|c|}{} & \\[-0,5ex]
        \multicolumn{2}{|c|}{Limiting Bessel kernels} & Case $\alpha=0$ \\
        \multicolumn{2}{|c|}{} & \\[-0,5ex]
        \hline
        & & \\[-0,5ex]
            $\mathbb{J}_{\alpha,I}(\zeta,\eta)$ &
            $\pi \zeta^{-\alpha+\frac{1}{2}}\eta^{-\alpha+\frac{1}{2}}
                \frac{J_{\alpha+\frac{1}{2}}(\pi\zeta)J_{\alpha-\frac{1}{2}}(\pi\eta)-
                J_{\alpha-\frac{1}{2}}(\pi\zeta)J_{\alpha+\frac{1}{2}}(\pi\eta)}{2(\zeta-\eta)}$ &
            $\frac{\sin\pi(\zeta-\eta)}{\pi(\zeta-\eta)}$ \\
        & & \\[-0,5ex]
        \hline
        & & \\[-0,5ex]
            $\mathbb{J}_{\alpha,II}^{+}(\zeta,\eta)$ &
            $\pi \zeta^{\alpha+\frac{1}{2}}\eta^{-\alpha+\frac{1}{2}}
                \frac{H_{\alpha+\frac{1}{2}}^{(1)}(\pi\zeta)J_{\alpha-\frac{1}{2}}(\pi\eta)-
                H_{\alpha-\frac{1}{2}}^{(1)}(\pi\zeta)J_{\alpha+\frac{1}{2}}(\pi\eta)}{4(\zeta-\eta)}$ &
            $-\frac{ie^{i\pi(\zeta-\eta)}}{2\pi(\zeta-\eta)}$ \\[4ex]
            $\mathbb{J}_{\alpha,II}^{-}(\zeta,\eta)$ &
            $-\pi \zeta^{\alpha+\frac{1}{2}}\eta^{-\alpha+\frac{1}{2}}
                \frac{H_{\alpha+\frac{1}{2}}^{(2)}(\pi\zeta)J_{\alpha-\frac{1}{2}}(\pi\eta)-
                H_{\alpha-\frac{1}{2}}^{(2)}(\pi\zeta)J_{\alpha+\frac{1}{2}}(\pi\eta)}{4(\zeta-\eta)}$ &
            $-\frac{ie^{-i\pi(\zeta-\eta)}}{2\pi(\zeta-\eta)}$ \\
        & & \\[-0,5ex]
        \hline
        & & \\[-0,5ex]
            $\mathbb{J}_{\alpha,III}^{+}(\zeta,\eta)$ &
            $\pi \zeta^{\alpha+\frac{1}{2}}\eta^{\alpha+\frac{1}{2}}
                \frac{H_{\alpha+\frac{1}{2}}^{(1)}(\pi\zeta)H_{\alpha-\frac{1}{2}}^{(1)}(\pi\eta)-
                H_{\alpha-\frac{1}{2}}^{(1)}(\pi\zeta)H_{\alpha+\frac{1}{2}}^{(1)}(\pi\eta)}{8(\zeta-\eta)}$ &
            0 \\[4ex]
            $\mathbb{J}_{\alpha,III}^{\pm}(\zeta,\eta)$ &
            $-\pi \zeta^{\alpha+\frac{1}{2}}\eta^{\alpha+\frac{1}{2}}
                \frac{H_{\alpha+\frac{1}{2}}^{(1)}(\pi\zeta)H_{\alpha-\frac{1}{2}}^{(2)}(\pi\eta)-
                H_{\alpha-\frac{1}{2}}^{(1)}(\pi\zeta)H_{\alpha+\frac{1}{2}}^{(2)}(\pi\eta)}{8(\zeta-\eta)}$ &
            $\frac{ie^{i\pi(\zeta-\eta)}}{2\pi(\zeta-\eta)}$ \\[4ex]
            $\mathbb{J}_{\alpha,III}^{-}(\zeta,\eta)$ &
            $\pi \zeta^{\alpha+\frac{1}{2}}\eta^{\alpha+\frac{1}{2}}
                \frac{H_{\alpha+\frac{1}{2}}^{(2)}(\pi\zeta)H_{\alpha-\frac{1}{2}}^{(2)}(\pi\eta)-
                H_{\alpha-\frac{1}{2}}^{(2)}(\pi\zeta)H_{\alpha+\frac{1}{2}}^{(2)}(\pi\eta)}{8(\zeta-\eta)}$ &
            0 \\[-0,5ex]
        & & \\
        \hline
    \end{tabular}
    \caption{Expressions for the limiting Bessel kernels. Here, $J_\nu$
        is the usual $J$-Bessel function of order
        $\nu$, and
        $H_{\nu}^{(1)}$ and $H_{\nu}^{(2)}$ are
        the Hankel functions of order $\nu$ of the first and second kind, respectively.
        The right column denotes the expressions
        in case $\alpha=0$.}
    \label{table:limiting-Bessel-kernels}
\end{center}
\end{table}

\medskip

Our results are given by the following three theorems. We use
$\mathbb{C}_+$ and $\mathbb{C}_-$ to denote the upper and lower
half-plane, respectively.

\begin{theorem}\label{theorem:WI}
    Fix $m\in\mathbb{Z}$, let $W_{I,n+m}$ be the kernel
    given in Table \ref{table:finite-kernels}, and let $\gamma_{j,n}>0$ be the
    leading coefficient of the $j$-th degree orthonormal polynomial with respect
    to $w_n$. For $\zeta,\eta\in\mathbb{C}$
    \begin{eqnarray}
        \nonumber
        \lefteqn{
        \gamma_{n+m-1,n}^2 \frac{1}{n\psi(0)}
            W_{I,n+m}\left(\frac{\zeta}{n\psi(0)},\frac{\eta}{n\psi(0)}\right)
            } \\[2ex]
        &&
            \qquad =\,
            \bigl(n\psi(0)\bigr)^{2\alpha}e^{nV(0)}
            \left(e^{\frac{V'(0)}{2\psi(0)}(\zeta+\eta)}\mathbb{J}_{\alpha,I}(\zeta,\eta)
                +O(1/n)\right),
    \end{eqnarray}
    as $n\to\infty$, where the Bessel kernel $\mathbb{J}_{\alpha,I}(\zeta,\eta)$
    is given in Table {\rm \ref{table:limiting-Bessel-kernels}}.
    The error term holds uniformly for $\zeta$ and $\eta$ in
    compact subsets of $\mathbb{C}$.
\end{theorem}

\begin{theorem}\label{theorem:WII}
    Fix $m\in\mathbb{Z}$, let $W_{II,n+m}$ be the kernel
    given in Table \ref{table:finite-kernels}, and let $\gamma_{j,n}>0$ be the
    leading coefficient of the $j$-th degree orthonormal polynomial with respect
    to $w_n$. Then the following holds.
    \begin{enumerate}
    \item[(a)]
        For $\zeta\in\mathbb{C}_+$ and $\eta\in\mathbb{C}$
        \begin{eqnarray}
            \nonumber
            \lefteqn{
            \gamma_{n+m-1,n}^2\frac{\zeta-\eta}{n\psi(0)}
            W_{II,n+m}\left(\frac{\zeta}{n\psi(0)},\frac{\eta}{n\psi(0)}\right)
            } \\[2ex]
            \label{theorem:WII:equation}
            && \qquad
            =\,
            (\zeta-\eta)e^{-\frac{V'(0)}{2\psi(0)}(\zeta-\eta)}\mathbb{J}_{\alpha,II}^+(\zeta,\eta)
                + O(1/n),
        \end{eqnarray}
        as $n\to\infty$, where the Bessel kernel $\mathbb{J}_{\alpha,II}^+(\zeta,\eta)$ is
        given in Table {\rm \ref{table:limiting-Bessel-kernels}}.
        The error term holds uniformly for $\zeta$ and $\eta$ in compact subsets of $\mathbb{C}_+$
        and $\mathbb{C}$, respectively.
    \item[(b)]
        For $\zeta\in\mathbb{C}_-$ and $\eta\in\mathbb{C}$
        \begin{eqnarray}
            \nonumber
            \lefteqn{
            \gamma_{n+m-1,n}^2\frac{\zeta-\eta}{n\psi(0)}W_{II,n+m}\left(\frac{\zeta}{n\psi(0)},
            \frac{\eta}{n\psi(0)}\right)
            } \\[2ex]
            && \qquad
            =\,
            (\zeta-\eta)e^{-\frac{V'(0)}{2\psi(0)}(\zeta-\eta)}\mathbb{J}_{\alpha,II}^-(\zeta,\eta)
                + O(1/n),
        \end{eqnarray}
        as $n\to\infty$, where the Bessel kernel $\mathbb{J}_{\alpha,II}^-(\zeta,\eta)$
        is given in Table {\rm \ref{table:limiting-Bessel-kernels}}.
        The error term holds uniformly for $\zeta$ and $\eta$ in compact subsets of $\mathbb{C}_-$
        and $\mathbb{C}$, respectively.
    \end{enumerate}
\end{theorem}

\begin{theorem}\label{theorem:WIII}
    Fix $m\in\mathbb{Z}$, let $W_{III,n+m}$ be the kernel
    given in Table \ref{table:finite-kernels}, and let $\gamma_{j,n}>0$ be the
    leading coefficient of the $j$-th degree orthonormal polynomial with respect
    to $w_n$. Then the following holds.
    \begin{enumerate}
    \item[(a)]
    For $\zeta,\eta\in\mathbb{C}_+$
    \begin{eqnarray}
        \nonumber
        \lefteqn{
        \gamma_{n+m-1,n}^2 \frac{1}{n\psi(0)}
            W_{III,n+m}\left(\frac{\zeta}{n\psi(0)},\frac{\eta}{n\psi(0)}\right)
            } \\[2ex]
        &&
            \qquad =\, \left(\frac{1}{n\psi(0)}\right)^{2\alpha}
            e^{-nV(0)}
            \left(e^{-\frac{V'(0)}{2\psi(0)}(\zeta+\eta)}\mathbb{J}_{\alpha,III}^+(\zeta,\eta)
                +O(1/n)\right),
    \end{eqnarray}
    as $n\to\infty$, where the Bessel kernel $\mathbb{J}_{\alpha,III}^+(\zeta,\eta)$
    is given in Table {\rm \ref{table:limiting-Bessel-kernels}}.
    The error term holds uniformly for $\zeta$ and $\eta$ in compact subsets
    of $\mathbb{C}_+$.
    \item[(b)]
    For $\zeta\in\mathbb{C}_+$ and $\eta\in\mathbb{C}_-$
    \begin{eqnarray}
        \nonumber
        \lefteqn{
        \gamma_{n+m-1,n}^2 \frac{1}{n\psi(0)}
            W_{III,n+m}\left(\frac{\zeta}{n\psi(0)},\frac{\eta}{n\psi(0)}\right)
            } \\[2ex]
        &&
            \qquad =\, \left(\frac{1}{n\psi(0)}\right)^{2\alpha}
            e^{-nV(0)}
            \left(e^{-\frac{V'(0)}{2\psi(0)}(\zeta+\eta)}\mathbb{J}_{\alpha,III}^\pm(\zeta,\eta)
                +O(1/n)\right),
    \end{eqnarray}
    as $n\to\infty$, where the Bessel kernel $\mathbb{J}_{\alpha,III}^\pm(\zeta,\eta)$ is
    given in Table {\rm \ref{table:limiting-Bessel-kernels}}.
    The error term holds uniformly for $\zeta$ and $\eta$ in compact subsets
    of $\mathbb{C}_+$ and $\mathbb{C}_-$, respectively.
    \item[(c)]
    For $\zeta,\eta\in\mathbb{C}_-$
    \begin{eqnarray}
        \nonumber
        \lefteqn{
        \gamma_{n+m-1,n}^2 \frac{1}{n\psi(0)}
            W_{III,n+m}\left(\frac{\zeta}{n\psi(0)},\frac{\eta}{n\psi(0)}\right)
            } \\[2ex]
        &&
            \qquad =\, \left(\frac{1}{n\psi(0)}\right)^{2\alpha}
            e^{-nV(0)}
            \left(e^{-\frac{V'(0)}{2\psi(0)}(\zeta+\eta)}\mathbb{J}_{\alpha,III}^-(\zeta,\eta)
                +O(1/n)\right),
    \end{eqnarray}
    as $n\to\infty$, where the Bessel kernel $\mathbb{J}_{\alpha,III}^-(\zeta,\eta)$
    is given in Table {\rm \ref{table:limiting-Bessel-kernels}}.
    The error term holds uniformly for $\zeta$ and $\eta$ in compact subsets
    of $\mathbb{C}_-$.
    \end{enumerate}
\end{theorem}

\begin{remark}
    In case $\alpha=0$ we can simplify the expressions for the
    limiting Bessel kernels, using the facts that, see
    \cite{AbramowitzStegun}
    \begin{eqnarray*}
        \lefteqn{
        J_{\frac{1}{2}}(z)=\sqrt{\frac{2}{\pi z}}\sin z,\qquad
        J_{-\frac{1}{2}}(z)=\sqrt{\frac{2}{\pi z}}\cos z,\qquad
        H^{(1)}_{\frac{1}{2}}(z)=-i\sqrt{\frac{2}{\pi z}}e^{iz},}
        \\[1ex]
        \lefteqn{
        H^{(1)}_{-\frac{1}{2}}(z)=\sqrt{\frac{2}{\pi
        z}}e^{iz},\qquad
        H^{(2)}_{\frac{1}{2}}(z)=i\sqrt{\frac{2}{\pi z}}e^{-iz},\qquad
        H^{(2)}_{-\frac{1}{2}}(z)=\sqrt{\frac{2}{\pi z}}e^{-iz}.
        }
    \end{eqnarray*}
    We then obtain the kernels given in the right column of Table
    \ref{table:limiting-Bessel-kernels}. This is in agreement with
    the results of Strahov and Fyodorov \cite{StrahovFyodorov}. Note however that in
    \cite{StrahovFyodorov} the second and the third kernel are
    multiplied with an extra factor $-2\pi i$.
\end{remark}

\begin{remark}
    As noted before, it has been shown by Strahov and Fyodorov
    \cite{StrahovFyodorov}, see also \cite{BaikDeiftStrahov}, that
    \[
        \left\langle \frac{\det(\frac{\eta}{n\psi(0)}-M)}{\det(\frac{\zeta}{n\psi(0)}-M)}
        \right\rangle_M = 2\pi i\gamma_{n-1,n}^2
        \frac{\zeta-\eta}{n\psi(0)}
        W_{II,n}\left(\frac{\zeta}{n\psi(0)},\frac{\eta}{n\psi(0)}\right).
    \]
    Then it follows from (\ref{theorem:WII:equation}), Table \ref{table:limiting-Bessel-kernels} and
    \cite[formula 9.1.3]{AbramowitzStegun}, that for
    $\zeta\in\mathbb{C}_+$,
    \[
        \left\langle \frac{\det(\frac{\zeta}{n\psi(0)}-M)}{\det(\frac{\zeta}{n\psi(0)}-M)}
        \right\rangle_M =
        \frac{\pi^2\zeta}{2}\left(J_{\alpha+\frac{1}{2}}(\pi\zeta)Y_{\alpha-\frac{1}{2}}(\pi\zeta)
        -J_{\alpha-\frac{1}{2}}(\pi\zeta)Y_{\alpha+\frac{1}{2}}(\pi\zeta)\right)+O(1/n),
    \]
    as $n\to\infty$, where $Y_\nu$ is the Bessel function of the second kind of order
    $\nu$. By \cite[formula
    9.1.16]{AbramowitzStegun}, the right hand side of
    this equation is $1+O(1/n)$, as it should be.
    Similarly we find the same result for $\zeta\in\mathbb{C}_-$.
\end{remark}

The proofs of these theorems are based on the characterization of
orthogonal polynomials with respect to the weight
(\ref{definition-weight}), together with their Cauchy transforms
via a $2\times2$ matrix RH problem for $Y$, due to Fokas, Its and
Kitaev \cite{FokasItsKitaev}, and on an application of the
Deift/Zhou steepest descent method \cite{DeiftZhou} for matrix RH
problems. See \cite{Deift,Kuijlaars} for an excellent exposition.
This technique was used before by Deift et al.\ \cite{DKMVZ2} to
establish universality for the local eigenvalue correlations in
unitary random matrix ensembles (\ref{unitary-ensemble-deift}) in
the bulk of the spectrum. Strahov and Fyodorov
\cite{StrahovFyodorov} used this method also to establish
universality for the three kernels $W_{I,n+m}, W_{II,n+m}$ and
$W_{III,n+m}$ in the bulk of the spectrum.

In a previous paper \cite{KV2} together with A.B.J.\ Kuijlaars,
the asymptotic analysis of the RH problem for $Y$, corresponding
to the weight (\ref{definition-weight}), has already been done. An
essential step in the analysis is the construction of the
parametrix near the origin, which gives us the behavior of $Y$
near the origin. Here, the Bessel functions come in. In
\cite{KV2}, the behavior of the first column of $Y$ (with the
orthogonal polynomials as entries) was determined near the origin
for positive (real) values, and used to establish universality for
the local eigenvalue correlations at the origin of the spectrum,
in terms of a Bessel kernel. Here, we determine the behavior of
the first column of $Y$, as well as the second column of $Y$ (with
the Cauchy transforms of orthogonal polynomials as entries) in a
{\em full} neighborhood of the origin, and use this in a similar
fashion to prove our results.

\medskip

The rest of the paper is organized as follows. In Section
\ref{section:the correspond RH problem} we give a short overview
of the asymptotic analysis of the corresponding RH problem for
$Y$. In Section \ref{section:behavior of Y near the origin} we
determine the behavior of $Y$ near the origin, in terms of Bessel
functions. This will be used in the last section to prove our
results.

\section{The corresponding RH problem}
    \label{section:the correspond RH problem}

In this section we recall the matrix RH problem for $Y$, due to
Fokas, Its and Kitaev \cite{FokasItsKitaev}, which characterizes
the orthogonal polynomials with respect to the weight
(\ref{definition-weight}), together with their Cauchy transforms.
We also give a short overview of the Deift/Zhou steepest descent
method \cite{Deift,DeiftZhou} to obtain the asymptotic behavior of
$Y$. For details we refer to \cite{DKMVZ2,KV2}, see also
\cite{Deift,DKMVZ1}.

\medskip

Our point of interest lies in the asymptotic behavior, as
$n\to\infty$, of the orthogonal polynomials $\pi_{n+m,n}$ of
degree $n+m$ with respect to the weight $w_n$, for any fixed
$m\in\mathbb{Z}$. So, in contrast to the RH problem in
\cite{DKMVZ2,KV2}, we have to modify the asymptotic condition at
infinity of the RH problem, and leave the jump condition
unchanged. However, this will not create any problems. We seek a
$2\times 2$ matrix valued function $Y=Y^{(n+m,n)}$ that satisfies
the following RH problem, cf.\
\cite{Deift,DKMVZ2,DKMVZ1,FokasItsKitaev,KV2}.

\subsubsection*{RH problem for \boldmath$Y$:}
\begin{enumerate}
    \item[(a)]
        $Y: \mathbb C \setminus \mathbb R \to \mathbb C^{2\times 2}$
        is  analytic.
    \item[(b)]
        $Y$ possesses continuous boundary values for $x \in \mathbb{R}\setminus\{0\}$
        denoted by $Y_{+}(x)$ and $Y_{-}(x)$, where $Y_{+}(x)$ and $Y_{-}(x)$
        denote the limiting values of $Y(z')$ as $z'$ approaches $x$ from
        above and below, respectively, and
        \begin{equation}\label{RHPYb}
            Y_+(x) = Y_-(x)
            \begin{pmatrix}
                1 & |x|^{2\alpha}e^{-nV(x)} \\
                0 & 1
            \end{pmatrix},
            \qquad\mbox{for $x \in \mathbb{R}\setminus\{0\}$.}
        \end{equation}
    \item[(c)]
        $Y$ has the following asymptotic behavior at infinity:
        \begin{equation} \label{RHPYc}
            Y(z)= (I+ O (1/z))
            \begin{pmatrix}
                z^{n+m} & 0 \\
                0 & z^{-(n+m)}
            \end{pmatrix}, \qquad \mbox{as $z\to\infty$.}
        \end{equation}
    \item[(d)]
        $Y$ has the following behavior near the origin:
        \begin{equation} \label{RHPYd}
            Y(z)=\left\{
            \begin{array}{cl}
                O\begin{pmatrix}
                    1 & |z|^{2\alpha} \\
                    1 & |z|^{2\alpha}
                \end{pmatrix}, &\mbox{if $\alpha< 0$,} \\[2ex]
                O\begin{pmatrix}
                    1 & 1 \\
                    1 & 1
                \end{pmatrix},
                &\mbox{if $\alpha>0$,}
            \end{array}\right.
        \end{equation}
        as $z \to 0$, $z \in \mathbb{C}\setminus\mathbb{R}$.
\end{enumerate}

\begin{remark}
    The $O$-terms in condition (d) of the RH problem are to be taken
    entrywise. So for example $Y(z)= O\begin{pmatrix} 1 & |z|^{2\alpha} \\
    1 & |z|^{2\alpha}  \end{pmatrix}$  means that
    $Y_{11}(z) = O(1)$, $Y_{12}(z) = O(|z|^{2\alpha})$, etc.
    This condition is used to control the
    behavior of $Y$ near the origin. In the following we will not
    go into detail about this condition, and refer to
    \cite{KMVV,Vanlessen} for details.
\end{remark}

The unique solution of the RH problem for $Y$, see
\cite{FokasItsKitaev} (for condition (d) we refer to \cite{KMVV}),
is then given by
\begin{equation}\label{RHPYsolution}
    Y(z) = Y^{(n+m,n)}(z)=
    \begin{pmatrix}
        \pi_{n+m,n}(z) &
            h_{n+m,n}(z) \\[2ex]
        -2\pi i \gamma_{n+m-1,n}^2 \pi_{n+m-1,n}(z) &
            -2\pi i\gamma_{n+m-1,n}^2 h_{n+m-1,n}(z)
    \end{pmatrix},
\end{equation}
for $z\in\mathbb{C}\setminus\mathbb{R}$, where $\pi_{j,n}$ is the
$j$-th degree monic orthogonal polynomial with respect to $w_n$,
where $\gamma_{j,n}$ is the leading coefficient of the $j$-th
degree orthonormal polynomial with respect to $w_n$, and where
$h_{j,n}$ is the Cauchy transform of $\pi_{j,n}$, see
(\ref{definition-Cauchy-transform}).

\begin{remark}
    The superscript $n+m$ in the notation $Y^{(n+m,n)}$ corresponds to the
    asymptotic condition (c) at infinity of the RH problem, which yields that the
    orthogonal polynomials in the solution (\ref{RHPYsolution}) of the RH
    problem have degree $n+m$ and $n+m-1$. The superscript $n$ corresponds to the
    jump condition (b), which  yields that the orthogonality is with
    respect to $w_n$.
\end{remark}

\begin{remark}
    We note that the first column of $Y$ contains the orthogonal
    polynomials, and the second column their Cauchy transforms.
    So, from Table \ref{table:finite-kernels} and (\ref{RHPYsolution}),
    the kernel $W_{I,n+m}$ depends only on the first column of
    $Y$, the kernel $W_{II,n+m}$ on both the first and the second column, and the
    kernel $W_{III,n+m}$ only on the second column, as follows:
    \begin{eqnarray}
        \label{remark: columnsY: eq1}
        \lefteqn{
        W_{I,n+m}(\zeta,\eta) = \frac{1}{\gamma_{n+m-1,n}^2} \frac{1}{-2\pi i (\zeta-\eta) }
            \det
            \begin{pmatrix}
                Y_{11}(\zeta) & Y_{11}(\eta) \\
                Y_{21}(\zeta) & Y_{21}(\eta)
            \end{pmatrix},
        } \\[2ex]
        \label{remark: columnsY: eq2}
        \lefteqn{
        W_{II,n+m}(\zeta,\eta) = \frac{1}{\gamma_{n+m-1,n}^2} \frac{1}{-2\pi i (\zeta-\eta)}
            \det
            \begin{pmatrix}
                Y_{12}(\zeta) & Y_{11}(\eta) \\
                Y_{22}(\zeta) & Y_{21}(\eta)
            \end{pmatrix},
        }
    \end{eqnarray}
    and
    \begin{equation} \label{remark: columnsY: eq3}
        W_{III,n+m}(\zeta,\eta) = \frac{1}{\gamma_{n+m-1,n}^2} \frac{1}{-2\pi i (\zeta-\eta)}
            \det
            \begin{pmatrix}
                Y_{12}(\zeta) & Y_{12}(\eta) \\
                Y_{22}(\zeta) & Y_{22}(\eta)
            \end{pmatrix}.
    \end{equation}
\end{remark}

\medskip

The asymptotic analysis of the RH problem for $Y$ includes a
series of transformations $Y\mapsto T\mapsto S\mapsto R$ to obtain
a RH problem for $R$ normalized at infinity (i.e. $R(z)\to I$ as
$n\to\infty$), and with jumps uniformly close to the identity
matrix, as $n\to\infty$. Then \cite{Deift,DKMVZ2,DKMVZ1}, $R$ is
also uniformly close to the identity matrix, and by unfolding the
series of transformations we obtain the asymptotic behavior of
$Y$.

Before we can give an overview of the series of transformations,
we need some properties of the equilibrium measure $\mu_V$ for
$V$. Here, we closely follow \cite{KV2}, see also
\cite{DKM,DKMVZ2}. The support of $\mu_V$ consists of a finite
union of intervals, say $\bigcup_{j=1}^{N+1}[b_{j-1},a_j]$, and we
define its interior as $J=\bigcup_{j=1}^{N+1}(b_{j-1},a_j)$. The
$N+1$ intervals of $J$ are referred to as the {\em bands}. The
density $\psi$ of the equilibrium measure is given by
\begin{equation}\label{definition-psi}
    \psi(x)=\frac{1}{2\pi i}R_+^{1/2}(x)h(x),\qquad \mbox{for $x\in J$,}
\end{equation}
with $h$ real analytic on $\mathbb{R}$, and where $R$ is the
$2(N+1)$-th degree monic polynomial with the endpoints $a_i,b_j$
of $J$ as zeros,
\begin{equation}
    R(z)=\prod_{j=1}^{N+1}(z-b_{j-1})(z-a_j).
\end{equation}
We use $R^{1/2}$ to denote the branch of $\sqrt R$ which behaves
like $z^{N+1}$ as $z\to\infty$, and which is defined and analytic
on $\mathbb{C}\setminus\bar J$. In (\ref{definition-psi}),
$R_+^{1/2}$ is used to denote the boundary value of $R^{1/2}$ on
$J$ from above. The equilibrium measure satisfies the
Euler-Lagrange variational conditions, which state that there
exists a constant $\ell\in\mathbb{R}$ such that
\begin{eqnarray}\label{Variational condition 1}
    \lefteqn{
    2\int\log|x-s|\psi(s)ds-V(x)= \ell,\qquad\mbox{for $x\in \bar
    J$,}} \\[1ex]
    \label{Variational condition 2}
    \lefteqn{
    2\int\log|x-s|\psi(s)ds-V(x)\leq \ell,\qquad\mbox{for $x\in\mathbb{R}\setminus\bar
    J$.}}
\end{eqnarray}
If the inequality in (\ref{Variational condition 2}) is strict for
every $x \in \mathbb R \setminus \bar J$, and if $h(x)\neq 0$ for
every $x \in \bar J$, then $V$ is called regular. Otherwise, there
are a finite number of points, called singular points of $V$, such
that $h$ vanishes there, i.e.\  a singular point in $\bar J$, or
such that we obtain equality in (\ref{Variational condition 2}),
i.e.\ a singular point in $\mathbb R \setminus \bar J$.

\medskip

Let $\sigma_3=\left(\begin{smallmatrix}1 & 0
\\ 0 & -1 \end{smallmatrix}\right)$ be the Pauli matrix. Following \cite{DKMVZ2}, see also \cite{KV2},
we define the $2\times 2$ matrix valued function
\begin{equation}\label{T-in-function-of-Y}
    T(z)=e^{-(n+m)\frac{\ell}{2}\sigma_3}Y(z)e^{(n+m)\frac{\ell}{2}\sigma_3}e^{-(n+m)g(z)\sigma_3},\qquad
    \mbox{for $z\in\mathbb{C}\setminus\mathbb{R}$,}
\end{equation}
where $\ell$ is the constant that appears in the Euler-Lagrange
variational conditions (\ref{Variational condition 1}) and
(\ref{Variational condition 2}), and where the scalar function $g$
is defined by
\begin{equation}\label{definition-g}
    g(z)=\int\log(z-s)\psi(s)ds,\qquad\mbox{for
    $z\in\mathbb{C}\setminus(-\infty,a_{N+1}]$.}
\end{equation}
Note the small difference in the definition
(\ref{T-in-function-of-Y}) of $T$ with its definition in
\cite{DKMVZ2,KV2}, which comes from the modified asymptotic
condition (c) of the RH problem for $Y$. For the case $m=0$, both
definitions agree. It is known \cite{DKMVZ2,KV2} that $T$ is
normalized at infinity and satisfies the jump relation
$T_+(x)=T_-(x)v^{(1)}(x)$ for $x\in\mathbb{R}\setminus\{0\}$,
where
\begin{equation}\label{definition-v1}
    v^{(1)}(x)=
    \left\{\begin{array}{l}
        \begin{pmatrix}
            e^{-(n+m)(g_+(x)-g_-(x))} & |x|^{2\alpha}e^{mV(x)}
            \\[1ex]
            0 & e^{(n+m)(g_+(x)-g_-(x))}
        \end{pmatrix},
            \quad x\in \bar{J} \setminus\{0\} \\[5ex]
        \begin{pmatrix}
            e^{-2\pi i(n+m)\Omega_j} & |x|^{2\alpha}e^{mV(x)}e^{(n+m)(g_+(x)+g_-(x)-V(x)-\ell)}
            \\[1ex]
            0 & e^{2\pi i(n+m)\Omega_j}
        \end{pmatrix}, \quad x\in(a_j,b_j),\ \ \ \ \ \\[5ex]
        \begin{pmatrix}
            1 & |x|^{2\alpha}e^{mV(x)} e^{(n+m)(g_+(x)+g_-(x)-V(x)-\ell)}
            \\[1ex]
            0 & 1
        \end{pmatrix},
            \quad \mbox{$x<b_0$ or $x>a_{N+1}$.}
    \end{array}\right.
\end{equation}
The constant $\Omega_j$ is the total $\mu_V$-mass of the $N+1-j$
largest bands.

\medskip

\begin{figure}
    \center{\resizebox{14cm}{!}{\includegraphics{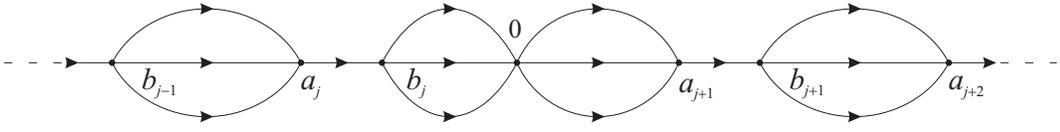}}
    \caption{Part of the contour $\Sigma$.
    }\label{figure: opening-of-the-lens}}
\end{figure}

The second transformation is referred to as the opening of the
lens. Define \cite{KV2} for every
$z\in\mathbb{C}\setminus\mathbb{R}$ lying in the region of
analyticity of $h$ the scalar function
\begin{equation}
    \phi(z)=\frac{1}{2}\int_{z}^{a_{N+1}}R^{1/2}(s)h(s)ds,
\end{equation}
where the path of integration does not cross the real axis. Then
\cite{KV2}, on the bands, $\phi$ is purely imaginary and satisfies
\begin{equation}\label{property-of-phi}
    2\phi_+(x)=-2\phi_-(x)=g_+(x)-g_-(x),\qquad \mbox{for $x\in
    J$,}
\end{equation}
so that $2\phi$ and $-2\phi$ provide analytic extensions of
$g_+-g_-$ into the upper half-plane and lower half-plane,
respectively. The opening of the lens is based on the
factorization of the jump matrix $v^{(1)}$ on the bands, see
(\ref{definition-v1}), into the following product of three
matrices, cf. \cite{KV2}
\begin{eqnarray}
    \nonumber
    \lefteqn{\begin{pmatrix}
        e^{-(n+m)(g_+(x)-g_-(x))} & |x|^{2\alpha}e^{mV(x)} \\[1ex]
        0 & e^{(n+m)(g_+(x)-g_-(x))}
        \end{pmatrix} \,=\,
        \begin{pmatrix}
        1 & 0 \\[1ex]
        |x|^{-2\alpha}e^{-mV(x)}e^{-2(n+m)\phi_-(x)} & 1
    \end{pmatrix}} \\[3ex]
    \nonumber
    && \qquad\qquad\times\,
    \begin{pmatrix}
        0 & |x|^{2\alpha}e^{mV(x)} \\[1ex]
        -|x|^{-2\alpha}e^{-mV(x)} & 0
    \end{pmatrix}
    \begin{pmatrix}
        1 & 0 \\[1ex]
        |x|^{-2\alpha}e^{-mV(x)}e^{-2(n+m)\phi_+(x)} & 1
    \end{pmatrix}.
\end{eqnarray}
We take an analytic continuation of the factor
$|x|^{2\alpha}e^{mV(x)}$ by defining for $z$ in the region of
analyticity of $V$,
\begin{equation}\label{definition-omega}
    \omega(z)=
    \left\{\begin{array}{ll}
        (-z)^{2\alpha}e^{mV(z)},&\qquad\mbox{if $\Re z < 0$,} \\[1ex]
        z^{2\alpha}e^{mV(z)},&\qquad\mbox{if $\Re z > 0$,}
    \end{array}\right.
\end{equation}
with principal branches of powers. We now open the lens. Let
$\Sigma$ be the lens shaped contour, as shown in Figure
\ref{figure: opening-of-the-lens}, going through the endpoints
$a_i,b_j$ of $J$, going trough the origin, and also going through
the singular points of $V$ in $J$. Of course we take the lens
shaped regions to lie within the region of analyticity of $\phi$
and $V$.

Define, cf.\ \cite{KV2}
\begin{equation} \label{S-in-function-of-T}
    S(z)=
    \left\{\begin{array}{cl}
        T(z), & \mbox{for $z$ outside the lens,} \\[2ex]
        T(z)
        \begin{pmatrix}
            1 & 0 \\
            -\omega(z)^{-1}e^{-2(n+m)\phi(z)} & 1
        \end{pmatrix}, & \mbox{for $z$ in the upper parts of the lens,} \\[3ex]
        T(z)
        \begin{pmatrix}
            1 & 0 \\
            \omega(z)^{-1}e^{-2(n+m)\phi(z)} & 1
        \end{pmatrix}, & \mbox{for $z$ in the lower parts of the lens.}
    \end{array} \right.
\end{equation}
As for the first transformation $Y\mapsto T$, there is small
difference in the definition (\ref{S-in-function-of-T}) for $S$
with its definition in \cite{KV2}, which comes from the modified
asymptotic condition (c) of the RH problem for $Y$. For the case
$m=0$, again both definitions agree. Then \cite{KV2}, the matrix
valued function $S$ is normalized at infinity and satisfies the
jump relation $S_+(z)=S_-(z)v^{(2)}(z)$ for $z\in\Sigma$, where
\begin{equation}
    v^{(2)}(z)=
    \left\{\begin{array}{l}
        \begin{pmatrix}
            1 & 0 \\[1ex]
            \omega(z)^{-1}e^{-2 (n+m)\phi(z)} & 1
        \end{pmatrix},
        \quad z\in\Sigma\cap\mathbb{C_\pm}, \\[5ex]
        \begin{pmatrix}
            0 & |z|^{2\alpha}e^{mV(z)} \\[1ex]
            -|z|^{-2\alpha}e^{-mV(z)} & 0
        \end{pmatrix},
        \quad z\in J\setminus\{0\}, \\[5ex]
        \begin{pmatrix}
            e^{-2\pi i(n+m)\Omega_j} & |z|^{2\alpha}e^{mV(z)}e^{(n+m)(g_+(z)+g_-(z)-V(z)-\ell)}
            \\[1ex]
            0 & e^{2\pi i(n+m)\Omega_j}
        \end{pmatrix}, \quad z\in(a_j,b_j) \\[5ex]
        \begin{pmatrix}
            1 & |z|^{2\alpha}e^{mV(z)} e^{(n+m)(g_+(z)+g_-(z)-V(z)-\ell)}
            \\[1ex]
            0 & 1
        \end{pmatrix},
        \quad \mbox{$z<b_0$ or $z>a_{N+1}$.}
    \end{array}\right.
\end{equation}

\medskip

For $z$ in a neighborhood of a regular point $x\in J$ we have,
cf.\ \cite{KV2},
\[
    \Re \phi(z)>0,\qquad \mbox{if $\Im z\neq 0$,}
\]
and for every regular point in $\mathbb{R}\setminus\bar J$ we have
from the Euler-Lagrange variational condition (\ref{Variational
condition 2}), cf.\ \cite{DKMVZ2}
\[
    g_+(x)+g_-(x)-V(x)-l<0, \qquad\mbox{for $x\in\mathbb{R}\setminus\bar J$.}
\]
So, we expect that the leading order asymptotics are determined by
a RH problem for $P^{(\infty)}$, normalized at infinity, that
satisfies the jump relation
$P^{(\infty)}_+(x)=P^{(\infty)}_-(x)v^{(\infty)}(x)$ for
$x\in(b_0,a_{N+1})$, where
\begin{equation}
    v^{(\infty)}(x)=
    \left\{\begin{array}{cl}
        \begin{pmatrix}
            0 & |x|^{2\alpha}e^{mV(x)} \\
            -|x|^{-2\alpha}e^{-mV(x)} & 0
        \end{pmatrix},
        & \mbox{for $x \in J\setminus\{0\}$,} \\[3ex]
        \begin{pmatrix}
            e^{-2\pi i(n+m)\Omega_j} & 0 \\
            0 & e^{2\pi i(n+m)\Omega_j}
        \end{pmatrix},
        & \mbox{for $x\in(a_j,b_j),j=1\ldots N$.}
    \end{array}\right.
\end{equation}
The solution of this RH problem is referred to as the parametrix
for the outside region, and is constructed using a Szeg\H{o}
function on multiple intervals associated to
$|x|^{2\alpha}e^{mV(x)}$, cf.\ \cite{KV2}, and using Riemann-Theta
functions, cf.\ \cite{DKMVZ2}, see also \cite{DIZ}. For our
purpose here, we do not need the explicit formulas for
$P^{(\infty)}$, and refer to \cite{DKMVZ2,KV2} for details.

\medskip

Before we can do the third transformation, we have to be careful
since the jump matrices for $S$ and $P^{(\infty)}$ are not
uniformly close to each other near 0, near the endpoints $a_i,b_j$
of $J$, and near the singular points of $V$. To solve this
problem, we surround these points by small non-overlapping disks,
say of radius $\delta>0$, and within each disk we construct a
parametrix $P$ satisfying the following local RH problem.

\subsubsection*{RH Problem for $P$ near $x_0$ where $x_0$ is 0, an endpoint of $J$,
or a singular point of $V$:}
\begin{enumerate}
\item[(a)]
    $P(z)$ is defined and analytic for $z \in
    \{ |z-x_0| < \delta_0 \}\setminus\Sigma$ for some $\delta_0 > \delta$.
\item[(b)]
    $P$ satisfies the same jump relations as $S$ does on
    $\Sigma \cap \{|z-x_0| < \delta\}$.
\item[(c)]
    There is $\kappa > 0$ such that,
    as $n \to \infty$,
    \begin{equation}\label{RHPPcbis}
        P(z) \left(P^{(\infty)}\right)^{-1}(z) = I +
        O ( 1/n^\kappa ),
        \qquad \mbox{uniformly for $|z-x_0| = \delta$.}
    \end{equation}
\item[(d)]
    $SP^{-1}$ has a removable singularity at $x_0$.
\end{enumerate}

For regular endpoints and the origin we can take $\kappa=1$ in
(\ref{RHPPcbis}). It is known that this local RH problem is
solvable for every $x_0$. For the endpoints of $J$ and the
singular points of $V$ we refer to \cite{DKMVZ2}, for the origin
we refer to \cite{KV2}. For our purpose here, it suffices to know
the explicit formula for the parametrix near the origin.

\medskip

We will now give the explicit formula for the parametrix $P$ near
the origin, see \cite[Section 5]{KV2} for details, see also
\cite[Section 4]{Vanlessen}. This is an essential step in the
asymptotic analysis of the RH problem since it allows us to
determine the behavior of $Y$ near the origin, which will be the
main tool to prove our results. Introduce the scalar function
\begin{equation}\label{definition-f}
    f(z)=
    \left\{
    \begin{array}{ll}
        i\phi(z)-i\phi_+(0), & \qquad\mbox{if $\Im z>0$,} \\
        -i\phi(z)-i\phi_+(0), & \qquad\mbox{if $\Im z<0$,}
    \end{array}\right.
\end{equation}
which is defined and analytic in a neighborhood of the origin. The
behavior of $f$ near the origin \cite[Section 5]{KV2} is given by
\begin{equation}\label{behavior-f}
    f(z)= \pi\psi(0)z+O(z^2),\qquad\mbox{as $z\to 0$.}
\end{equation}
Let $U_\delta$ be the disk with radius $\delta$ around the origin,
with $\delta>0$ sufficiently small such that $U_\delta$ lies in
the region of analyticity of $\phi$ and $V$. Since
$f'(0)=\pi\psi(0)> 0$ we can choose $\delta$ also sufficiently
small such that $f$ is a conformal mapping on $U_\delta$ onto a
convex neighborhood of 0. We have that $f(x)$ is real and positive
(negative) for $x\in U_\delta$ positive (negative).

\begin{figure}
    \center{\resizebox{13cm}{!}{\includegraphics{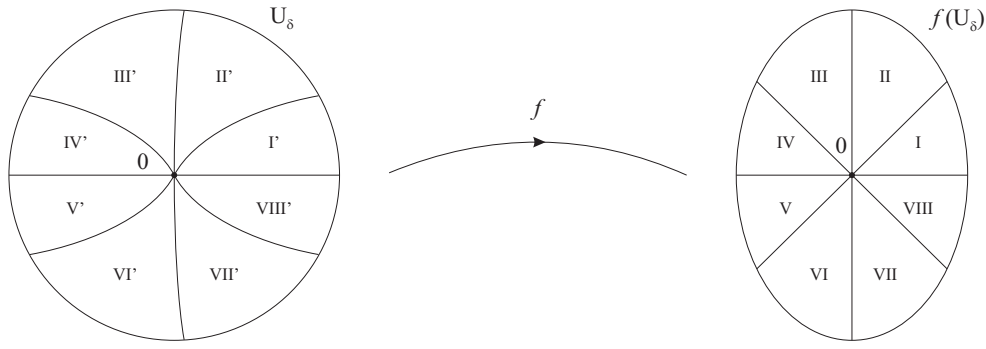}}
    \caption{Decomposition of $U_\delta$ and $f(U_\delta)$ into eight regions.}
    \label{figure:conformal-mapping-f}}
\end{figure}

Decompose $f(U_\delta)$ into eight regions I--VIII, as shown in
the right of Figure \ref{figure:conformal-mapping-f}, divided by
eight straight rays
\[
    \Gamma_j=\{ \zeta \in \mathbb C \mid \arg \zeta = (j-1)\frac{\pi}{4}
    \}, \qquad \mbox{j=1,\ldots ,8.}
\]
This in turn divides the disk $U_\delta$ into eight regions
I'--VIII' as the pre-images under $f$ of I--VIII, as shown in the
left of Figure \ref{figure:conformal-mapping-f}. Sector I' and IV'
correspond to the right and left upper part of the lens inside
$U_\delta$, respectively, sector V' and VIII' to the left and
right lower part of the lens inside $U_\delta$, respectively.

Let $\Psi_\alpha$ be the piecewise analytic matrix valued function
\cite[Section 4]{Vanlessen}, see also \cite[Section 5]{KV2}, that
satisfies the jump relation
$\Psi_{\alpha,+}(\zeta)=\Psi_{\alpha,-}(\zeta)v_\alpha(\zeta)$ for
$\zeta\in\bigcup \Gamma_j$, where
\[
    v_\alpha(\zeta)=\left\{\begin{array}{cl}
        \begin{pmatrix}
            0 & 1 \\
            -1 & 0
        \end{pmatrix},
            & \qquad\mbox{for $\zeta \in \Gamma_1\cup\Gamma_5$,} \\[3ex]
        \begin{pmatrix}
            1 & 0 \\
            e^{-2\pi i \alpha}  & 1
        \end{pmatrix},
            & \qquad\mbox{for $\zeta \in \Gamma_2\cup\Gamma_6$,} \\[3ex]
        e^{\pi i \alpha\sigma_3},
            & \qquad\mbox{for $\zeta \in \Gamma_3\cup\Gamma_7$,} \\[2ex]
        \begin{pmatrix}
            1 & 0 \\
            e^{2\pi i \alpha}  & 1
        \end{pmatrix},
            & \qquad\mbox{for $\zeta \in \Gamma_4\cup\Gamma_8$,}
        \end{array}\right.
\]
and that has the following behavior near the origin,
\[
    \Psi_\alpha(\zeta)=
        O\begin{pmatrix}
            |\zeta|^\alpha & |\zeta|^\alpha \\
            |\zeta|^\alpha & |\zeta|^\alpha
        \end{pmatrix},
        \qquad \mbox{as $\zeta\to 0$,}
\]
if $\alpha<0$, and
\[
        \Psi_\alpha(\zeta)=
        \left\{\begin{array}{cl}
            O\begin{pmatrix}
                |\zeta|^\alpha & |\zeta|^{-\alpha} \\
                |\zeta|^\alpha & |\zeta|^{-\alpha}
            \end{pmatrix},
            & \mbox{as $\zeta\to 0$ for $\frac{\pi}{4}<|\arg\zeta|< \frac{3\pi}{4}$,} \\[3ex]
            O\begin{pmatrix}
                |\zeta|^{-\alpha} & |\zeta|^{-\alpha}\\
                |\zeta|^{-\alpha} & |\zeta|^{-\alpha}
            \end{pmatrix},
            & \mbox{as $\zeta\to 0$ for $0<|\arg\zeta|<\frac{\pi}{4}$ and $\frac{3\pi}{4}<|\arg\zeta|
            <\pi$,}
        \end{array}\right.
\]
if $\alpha>0$. The behavior of $\Psi_\alpha$ near the origin will
ensure that part (d) of the RH problem for $P$ is satisfied, see
\cite{KV2,Vanlessen} for details. The matrix valued function
$\Psi_\alpha$ is constructed out of Bessel functions of order
$\alpha\pm\frac{1}{2}$, and its explicit formula for
$0<\arg\zeta<\frac{\pi}{4}$ is given by
\begin{equation}\label{PsiAlphaI}
    \Psi_\alpha(\zeta)=\frac{1}{2}\sqrt\pi\zeta^{1/2}
    \begin{pmatrix}
        H_{\alpha+\frac{1}{2}}^{(2)}(\zeta) &
            -iH_{\alpha+\frac{1}{2}}^{(1)}(\zeta) \\[2ex]
        H_{\alpha-\frac{1}{2}}^{(2)}(\zeta) &
            -iH_{\alpha-\frac{1}{2}}^{(1)}(\zeta)
    \end{pmatrix}e^{-(\alpha+\frac{1}{4})\pi
    i\sigma_3}.
\end{equation}
For $\frac{\pi}{4}<\arg\zeta<\frac{\pi}{2}$ it is given by
\begin{equation}\label{PsiAlphaII}
    \Psi_{\alpha}(\zeta)=
    \begin{pmatrix}
        \sqrt\pi\zeta^{1/2}I_{\alpha+\frac{1}{2}}(\zeta e^{-\frac{\pi i}{2}}) &
            -\frac{1}{\sqrt\pi}\zeta^{1/2}K_{\alpha+\frac{1}{2}}(\zeta e^{-\frac{\pi i}{2}}) \\[1ex]
        -i\sqrt\pi\zeta^{1/2}I_{\alpha-\frac{1}{2}}(\zeta e^{-\frac{\pi i}{2}}) &
            -\frac{i}{\sqrt\pi}\zeta^{1/2}K_{\alpha-\frac{1}{2}}(\zeta e^{-\frac{\pi i}{2}})
    \end{pmatrix}
    e^{-\frac{1}{2}\pi i\alpha\sigma_3},
\end{equation}
where $I_\nu$ and $K_\nu$ are the modified Bessel functions of
order $\nu$. See \cite[Section 4]{Vanlessen} for the explicit
expressions of $\Psi_\alpha$ in the other sectors of the complex
plane. Also define the piecewise analytic function $W$ by
\begin{equation}\label{definition-W}
    W(z) =
        \left\{\begin{array}{ll}
            z^\alpha e^{m\frac{V(z)}{2}} , & \qquad \mbox{if $z\in$ III',IV',V',VI',} \\[1ex]
            (-z)^\alpha e^{m\frac{V(z)}{2}}, & \qquad \mbox{if $z\in$ I',II',VII',VIII'.}
        \end{array}\right.
\end{equation}
And finally, define the following matrix valued function, analytic
in a neighborhood of the disk $U_\delta$,
\begin{equation}\label{definition-En}
    E_{n+m,n}(z)=E(z) e^{(n+m)\phi_+(0)\sigma_3} e^{-\frac{\pi i}{4}\sigma_3} \frac{1}{\sqrt 2}
    \begin{pmatrix}
        1 & i \\
        i & 1
    \end{pmatrix},
\end{equation}
where the matrix valued function $E$ is given by
\cite[(5.27)--(5.30)]{KV2}.

Then, cf.\ \cite[Section 5]{KV2}, the parametrix near the origin
is defined by
\begin{equation}\label{definition-P}
    P(z)=E_{n+m,n}(z)
        \Psi_\alpha\bigl((n+m)f(z)\bigr)W(z)^{-\sigma_3}e^{-(n+m)\phi(z)\sigma_3}.
\end{equation}

\begin{remark}
In contrast to \cite[Section 5]{KV2}, we evaluate the matrix
valued function $\Psi_\alpha$ in $(n+m)f(z)$ instead of in
$nf(z)$. This comes from the fact that, in order that the matching
condition (c) of the RH problem for $P$ is satisfied, we need to
cancel out the factor $e^{-(n+m)\phi(z)\sigma_3}$ instead of
$e^{-n\phi(z)}$. This follows in essence from the modified
asymptotic condition (c) of the RH problem for $Y$. For the case
$m=0$, the definition (\ref{definition-P}) of the parametrix $P$
near the origin agrees with its definition in \cite[Section
5]{KV2}.
\end{remark}

\medskip

\begin{figure}
    \center{\resizebox{14cm}{!}{\includegraphics{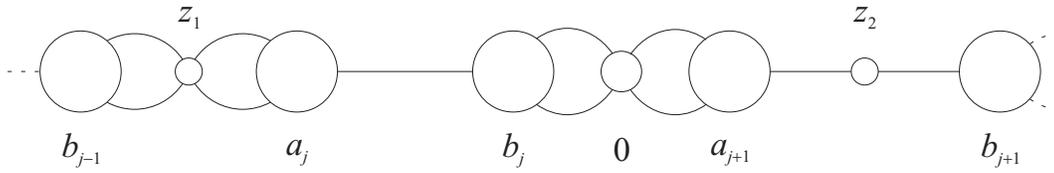}}
    \caption{Part of the contour $\Sigma_R$. The points $z_1$ and $z_2$
    are singular points of $V$.}\label{figure:system-of-contours-R}}
\end{figure}

Now, we have all the ingredients to give the third transformation.
Define \cite{DKMVZ2,KV2} the $2\times 2$ matrix valued function
$R$ as
\begin{equation}\label{R-in-function-of-S}
    R(z)=\left\{
    \begin{array}{ll}
        S(z)\left(P^{(\infty)}\right)^{-1}(z),&
            \qquad \mbox{for $z$ outside the disks,} \\[1ex]
        S(z)P^{-1}(z), & \qquad \mbox{for $z$ inside the disks.}
    \end{array}\right.
\end{equation}
Then \cite{DKMVZ2,KV2}, $R$ is normalized at infinity, and
analytic on the entire plane except for jumps on the reduced
system of contours $\Sigma_R$, as shown in Figure
\ref{figure:system-of-contours-R}, and except for possible
isolated singularities at the endpoints $a_i,b_j$ of $J$, at the
singularities of $V$ and at 0. However, from condition (d) of the
RH problem for $P$, these singularities are removable, so that $R$
is analytic on $\mathbb{C}\setminus\Sigma_R$. It is known
\cite{DKMVZ2,KV2} that the jumps of $R$ on $\Sigma_R$ are
uniformly close to the identity matrix as $n\to\infty$. This
implies \cite{DKMVZ2}, see also \cite{Deift,DKMVZ1}
\begin{equation}
    R(z)=I+O(1/n^\kappa),\qquad\mbox{as $n\to\infty$,}
\end{equation}
uniformly for $z\in\mathbb{C}\setminus\Sigma_R$, where $\kappa$ is
the constant that appears in the matching condition (c) of the RH
problem for $P$. By tracing back the steps $Y\mapsto T\mapsto
S\mapsto R$ we obtain the asymptotic behavior of $Y$ in all
regions of the complex plane, as $n\to\infty$.

\section{Behavior of $Y$ near the origin}
    \label{section:behavior of Y near the origin}

In this section we unravel, as in \cite[Lemma 7.1]{KV2}, the
series of transformations $Y\mapsto T\mapsto S\mapsto R$, see
Section \ref{section:the correspond RH problem}, to determine the
behavior of the first and the second column of $Y$ inside the disk
$U_\delta$. This behavior will be the main tool to prove our
results. Note that the second column of $Y$ has jumps on the real
axis, see (\ref{RHPYb}). So, for the behavior of the second column
of $Y$ inside the disk $U_\delta$ we have to distinguish between
the upper and lower parts of $U_\delta$.

\medskip

For notational convenience we introduce the $2\times 2$ matrix
valued function, cf.\ \cite[Lemma 7.1]{KV2}
\begin{equation}\label{definition-M}
    M(z)=M_{n+m,n}(z)=R(z)E_{n+m,n}(z),\qquad \mbox{for $z\in U_\delta$,}
\end{equation}
where $E_{n+m,n}$ is given by (\ref{definition-En}). For the case
$m=0$, the $M$-matrix defined by (\ref{definition-M}) corresponds
to the $M$-matrix in \cite[Lemma 7.1]{KV2}. It is known that $M$
is analytic on $U_\delta$, that each entry of $M$ is uniformly
bounded in $U_\delta$ as $n\to\infty$, and that $\det M\equiv 1$,
cf.\ \cite[Lemma 7.1]{KV2}.

We also need the following lemma.

\begin{lemma}\label{lemma:behavior-Y}
    For $z\in U_\delta$,
    \begin{equation}\label{lemma:behavior-Y:equation}
        2g(z)-2\phi(z)-\ell=V(z).
    \end{equation}
\end{lemma}

\begin{proof}
    Let $H(z)=2g(z)-2\phi(z)-\ell-V(z)$, which is defined and
    analytic for $z\in U_\delta\setminus\mathbb{R}$. For
    $x\in(-\delta,\delta)\subset J$ we have by (\ref{property-of-phi})
    \begin{equation}\label{lemma:behavior-Y:eq1}
        H_+(x)=H_-(x)=g_+(x)+g_-(x)-\ell-V(x),
    \end{equation}
    so that $H$ is analytic in the entire disk $U_\delta$. For
    $x\in(-\delta,\delta)$ we have by (\ref{definition-g})
    \[
        g_+(x)+g_-(x)=2\int\log|x-s|\psi(s)ds.
    \]
    Inserting this into (\ref{lemma:behavior-Y:eq1}) and using the
    Euler-Lagrange variational condition (\ref{Variational condition
    1}), we have that $H(x)=0$ for
    $x\in(-\delta,\delta)$. This implies from the
    uniqueness principle
    that $H\equiv 0$ on $U_\delta$, which proves the
    lemma.
\end{proof}

First, the behavior of the first column of $Y$ inside the disk
$U_\delta$ is given by the following theorem.

\begin{theorem}\label{theorem:first-column-Y}
    Fix $m\in\mathbb{Z}$. For $z\in U_\delta$ and $n$ sufficiently large,
    the first column of $Y=Y^{(n+m,n)}$ is given by
    \begin{eqnarray}
        \nonumber
        \lefteqn{
        \begin{pmatrix}
            Y_{11}(z) \\
            Y_{21}(z)
        \end{pmatrix}
        =
        z^{-\alpha} e^{n\frac{V(z)}{2}} \sqrt\pi e^{-\frac{\pi
        i}{4}} e^{(n+m)\frac{\ell}{2}\sigma_3}M(z)} \\
        \label{theorem:first-column-Y:equation}
        && \qquad\qquad\qquad\qquad\times\,
        \begin{pmatrix}
            \bigl((n+m)f(z)\bigr)^{1/2} J_{\alpha+\frac{1}{2}}\bigl((n+m)f(z)\bigr) \\[1ex]
            \bigl((n+m)f(z)\bigr)^{1/2} J_{\alpha-\frac{1}{2}}\bigl((n+m)f(z)\bigr)
        \end{pmatrix},
    \end{eqnarray}
    Here, $J_\nu$ is the $J$-Bessel
    function of order $\nu$, $f$ is given by {\rm (\ref{definition-f})},
    and $M$ is given by {\rm (\ref{definition-M})}.
\end{theorem}

\begin{proof}
    Let $z$ be in sector I' of the disk $U_\delta$, see Figure \ref{figure:conformal-mapping-f}.
    Unfolding the series of transformations $Y\mapsto T\mapsto S\mapsto R$ we obtain by
    (\ref{T-in-function-of-Y}), (\ref{S-in-function-of-T}),
    (\ref{definition-P}) and (\ref{R-in-function-of-S})
    \begin{eqnarray}
        \nonumber
        Y(z) &=&
        e^{(n+m)\frac{\ell}{2}\sigma_3}R(z)E_{n+m,n}(z)\Psi_\alpha\bigl((n+m)f(z)\bigr)W(z)^{-\sigma_3}
        \\[1ex]
        \label{theorem:first-column-Y:eq1a}
        && \qquad\times\,
        e^{-(n+m)\phi(z)\sigma_3}
        \begin{pmatrix}
            1 & 0 \\
            \omega(z)^{-1}e^{-2(n+m)\phi(z)} & 1
        \end{pmatrix}
        e^{-(n+m)\frac{\ell}{2}\sigma_3}e^{(n+m)g(z)\sigma_3}.
    \end{eqnarray}
    Note that $\omega(z)=z^{2\alpha}e^{mV(z)}$, see
    (\ref{definition-omega}), and that
    $W(z)=(-z)^\alpha e^{m\frac{V(z)}{2}}=z^\alpha e^{-\pi i\alpha}e^{m\frac{V(z)}{2}}$
    , see (\ref{definition-W}). Inserting this into (\ref{theorem:first-column-Y:eq1a})
    and using (\ref{definition-M}) and (\ref{lemma:behavior-Y:equation}),
    the first column of $Y$ is then given by
    \begin{equation}\label{theorem:first-column-Y:eq1}
        \begin{pmatrix}
            Y_{11}(z) \\
            Y_{21}(z)
        \end{pmatrix}
        =
        z^{-\alpha}e^{n\frac{V(z)}{2}}e^{(n+m)\frac{\ell}{2}\sigma_3}M(z)
        \Psi_\alpha\bigl((n+m)f(z)\bigr)e^{\pi i\alpha\sigma_3}
        \begin{pmatrix}
            1 \\
            1
        \end{pmatrix}.
    \end{equation}
    Since $f(z)$ is in sector I of $f(U_\delta)$, see Figure
    \ref{figure:conformal-mapping-f}, we have for $n$ sufficiently large (namely
    $n+m>0$) that $0<\arg (n+m)f(z)<\pi/4$. So, we have to use (\ref{PsiAlphaI})
    to evaluate $\Psi_\alpha\bigl((n+m)f(z)\bigr)$. From (\ref{theorem:first-column-Y:eq1})
    and \cite[formulas 9.1.3 and 9.1.4]{AbramowitzStegun}, which connect the Hankel functions
    of the first and second kind with the ordinary $J$-Bessel
    functions, we then establish (\ref{theorem:first-column-Y:equation}) in sector I' of $U_\delta$.

    Now, let $z$ be in sector II' of $U_\delta$. Similarly as in sector I', we
    obtain by
    (\ref{T-in-function-of-Y}), (\ref{S-in-function-of-T}),
    (\ref{definition-P}) and (\ref{R-in-function-of-S})
    \begin{eqnarray}
        \nonumber
            Y(z)&=& e^{(n+m)\frac{\ell}{2}\sigma_3} R(z) E_{n+m,n}(z) \Psi_\alpha\bigl((n+m)f(z)\bigr)
        \\[1ex]
        \nonumber
        && \qquad \times\,
            W(z)^{-\sigma_3}
            e^{-(n+m)\phi(z)\sigma_3}
            e^{-(n+m)\frac{\ell}{2}\sigma_3}e^{(n+m)g(z)\sigma_3}.
    \end{eqnarray}
    Since $W(z)=z^\alpha e^{-\pi i\alpha}e^{m\frac{V(z)}{2}}$
    , see (\ref{definition-W}), and using
    (\ref{definition-M}) and (\ref{lemma:behavior-Y:equation}),
    the first column of $Y$ is then given by
    \begin{equation}\label{theorem:first-column-Y:eq2}
        \begin{pmatrix}
            Y_{11}(z) \\
            Y_{21}(z)
        \end{pmatrix}
        =
        z^{-\alpha}e^{n\frac{V(z)}{2}}e^{(n+m)\frac{\ell}{2}\sigma_3}
        M(z) \Psi_\alpha\bigl((n+m)f(z)\bigr) e^{\pi i\alpha\sigma_3}
        \begin{pmatrix}
            1 \\
            0
        \end{pmatrix}.
    \end{equation}
    Since $\pi/4<\arg(n+m)f(z)<\pi/2$ for $n$ sufficiently large, we have to use
    (\ref{PsiAlphaII}) to
    evaluate $\Psi_\alpha\bigl((n+m)f(z)\bigr)$. This implies, using
    \cite[formula 9.6.3]{AbramowitzStegun}, which connects the modified
    Bessel function $I_{\alpha\pm\frac{1}{2}}$ with the Bessel function
    $J_{\alpha\pm\frac{1}{2}}$, that
    \begin{eqnarray}
        \nonumber
        \lefteqn{
        \Psi_\alpha\bigl((n+m)f(z)\bigr)e^{\pi i\alpha\sigma_3}
        \begin{pmatrix}
            1 \\
            0
        \end{pmatrix}} \\[1ex]
        &=&
        \sqrt\pi e^{\frac{\pi i\alpha}{2}}
        \begin{pmatrix}
            \bigl((n+m)f(z)\bigr)^{1/2} I_{\alpha+\frac{1}{2}}\bigl((n+m)f(z)e^{-\frac{\pi i}{2}}\bigr) \\[1ex]
            -i \bigl((n+m)f(z)\bigr)^{1/2} I_{\alpha-\frac{1}{2}}\bigl((n+m)f(z)e^{-\frac{\pi i}{2}}\bigr)
        \end{pmatrix} \\[2ex]
        \nonumber
        &=&
        \sqrt\pi e^{-\frac{\pi i}{4}}
        \begin{pmatrix}
            \bigl((n+m)f(z)\bigr)^{1/2} J_{\alpha+\frac{1}{2}}\bigl((n+m)f(z)\bigr) \\[1ex]
            \bigl((n+m)f(z)\bigr)^{1/2} J_{\alpha-\frac{1}{2}}\bigl((n+m)f(z)\bigr)
        \end{pmatrix}.
    \end{eqnarray}
    Inserting this into (\ref{theorem:first-column-Y:eq2}), we establish
    (\ref{theorem:first-column-Y:equation}) also in sector II' of $U_\delta$.

    In the other sectors of the disk $U_\delta$ the calculations are
    similar, and are left as an easy exercise for the careful reader.
\end{proof}

\begin{remark}
    For the case $m=0$, this theorem agrees with
    \cite[Lemma 7.1]{KV2}.
\end{remark}

\begin{remark}
    It is not quite clear from (\ref{theorem:first-column-Y:equation}) that the first column of $Y$ is
    analytic in the entire disk $U_\delta$, which must be the case since
    it has polynomials as entries, see (\ref{RHPYsolution}).
    Obviously, it is analytic on $U_\delta\setminus (-\delta,0]$.
    From \cite[formula 9.1.35]{AbramowitzStegun} we have for $x\in(-\delta,0)$
    \begin{eqnarray}\label{remark:first-column-Y:equation}
        \nonumber
        \lefteqn{
        \begin{pmatrix}
            Y_{11,+}(x) \\
            Y_{21,+}(x)
        \end{pmatrix}
        =
        \begin{pmatrix}
            Y_{11,-}(x) \\
            Y_{21,-}(x)
        \end{pmatrix}
        =\, |x|^{-\alpha}e^{n\frac{V(x)}{2}}
        \sqrt\pi e^{-\frac{\pi
        i}{4}}e^{(n+m)\frac{\ell}{2}\sigma_3}M(x)} \\[1ex]
        && \qquad\qquad\qquad\qquad \times\,
        \begin{pmatrix}
            -\bigl(-(n+m)f(x)\bigr)^{1/2}J_{\alpha+\frac{1}{2}}\bigl(-(n+m)f(x)\bigr) \\[1ex]
            \bigl(-(n+m)f(x)\bigr)^{1/2}J_{\alpha-\frac{1}{2}}\bigl(-(n+m)f(x)\bigr)
        \end{pmatrix}.
    \end{eqnarray}
    So, the first column of $Y$ is analytic in the entire disk $U_\delta$
    except for a possible isolated singularity at the origin. Since
    $J_{\alpha\pm\frac{1}{2}}(z)=O(z^{\alpha\pm\frac{1}{2}})$ as
    $z\to 0$, see \cite[formula 9.1.10]{AbramowitzStegun}
    this singularity is removable, which implies that the first
    column of $Y$ is analytic in the entire disk.
\end{remark}

Next, the behavior of the second column of $Y$ in the upper part
of the disk $U_\delta$ is given by the following theorem.

\begin{theorem}\label{theorem:second-column-Y-upper-part}
        Fix $m\in\mathbb{Z}$. For $z\in U_\delta\cap\mathbb{C}_+$ and $n$ sufficiently large, the second
        column of $Y=Y^{(n+m,n)}$ is given by
        \begin{eqnarray}
            \nonumber
            \lefteqn{
            \begin{pmatrix}
                Y_{12}(z) \\[1ex]
                Y_{22}(z)
            \end{pmatrix}
            =
            z^\alpha e^{-n\frac{V(z)}{2}} \frac{1}{2} \sqrt\pi e^{-\frac{\pi
            i}{4}} e^{(n+m)\frac{\ell}{2}\sigma_3} M(z)} \\
            \label{theorem:second-column-Y-upper-part:equation}
            && \qquad\qquad\qquad\qquad \times\,
            \begin{pmatrix}
                \bigl((n+m)f(z)\bigr)^{1/2} H_{\alpha+\frac{1}{2}}^{(1)}\bigl((n+m)f(z)\bigr) \\[2ex]
                \bigl((n+m)f(z)\bigr)^{1/2} H_{\alpha-\frac{1}{2}}^{(1)}\bigl((n+m)f(z)\bigr)
            \end{pmatrix}.
        \end{eqnarray}
        Here, $H^{(1)}_\nu$ is the Hankel
        function of the first kind of order $\nu$, $f$ is given by {\rm (\ref{definition-f})},
        and $M$ is given by {\rm (\ref{definition-M})}.
\end{theorem}

\begin{proof}
    Let $z$ be in sector I' of $U_\delta$. Instead of
    (\ref{theorem:first-column-Y:eq1}) we get for the second column of $Y$
    \begin{equation}\label{theorem:second-column-Y-upper-part:eq1}
        \begin{pmatrix}
            Y_{12}(z) \\
            Y_{22}(z)
        \end{pmatrix}
        =
        z^\alpha e^{-n\frac{V(z)}{2}}
        e^{(n+m)\frac{\ell}{2}\sigma_3}M(z)\Psi_\alpha\bigl((n+m)f(z)\bigr)e^{\pi
        i\alpha\sigma_3}
        \begin{pmatrix}
            0 \\
            1
        \end{pmatrix}.
    \end{equation}
    Since $0<\arg(n+m)f(z)<\pi/4$ for $n$ sufficiently large, we have to use (\ref{PsiAlphaI}) to evaluate
    $\Psi_\alpha\bigl((n+m)f(z)\bigr)$. Inserting this into
    (\ref{theorem:second-column-Y-upper-part:eq1}), we obtain
    (\ref{theorem:second-column-Y-upper-part:equation}) for this choice of $z$.

    Now, let $z$ be in sector II' of $U_\delta$. Instead of (\ref{theorem:first-column-Y:eq2})
    the second column of $Y$ is given by
    \begin{equation}\label{theorem:second-column-Y-upper-part:eq2}
        \begin{pmatrix}
            Y_{12}(z) \\
            Y_{22}(z)
        \end{pmatrix}
        =
        z^\alpha
        e^{-n\frac{V(z)}{2}}e^{(n+m)\frac{\ell}{2}\sigma_3}M(z)
        \Psi_\alpha\bigl((n+m)f(z)\bigr)e^{\pi i\alpha\sigma_3}
        \begin{pmatrix}
            0 \\
            1
        \end{pmatrix}.
    \end{equation}
    Since $\pi/4<\arg(n+m)f(z)<\pi/2$ for $n$ sufficiently large, we have to use (\ref{PsiAlphaII}) to evaluate
    $\Psi_\alpha\bigl((n+m)f(z)\bigr)$. From
    \cite[formula 9.6.4]{AbramowitzStegun},
    which connects the modified Bessel function
    $K_{\alpha\pm\frac{1}{2}}$ with the Hankel function
    $H_{\alpha\pm\frac{1}{2}}^{(1)}$ of the first kind, we then have
    \begin{eqnarray}
        \nonumber
        \lefteqn{
        \Psi_\alpha\bigl((n+m)f(z)\bigr) e^{\pi i\alpha\sigma_3}
        \begin{pmatrix}
            0 \\
            1
        \end{pmatrix}} \\[1ex]
        \nonumber
        &=&
        -\frac{1}{\sqrt\pi}e^{-\frac{\pi i\alpha}{2}}
        \begin{pmatrix}
            \bigl((n+m)f(z)\bigr)^{1/2} K_{\alpha+\frac{1}{2}}\bigl((n+m)f(z)e^{-\frac{\pi i}{2}}\bigr) \\[2ex]
            i \bigl((n+m)f(z)\bigr)^{1/2} K_{\alpha-\frac{1}{2}}\bigl((n+m)f(z)e^{-\frac{\pi i}{2}}\bigr)
        \end{pmatrix} \\[2ex]
        \nonumber
        &=&
        \frac{1}{2}\sqrt\pi e^{-\frac{\pi i}{4}}
        \begin{pmatrix}
            \bigl((n+m)f(z)\bigr)^{1/2} H_{\alpha+\frac{1}{2}}^{(1)}\bigl((n+m)f(z)\bigr) \\[2ex]
            \bigl((n+m)f(z)\bigr)^{1/2} H_{\alpha-\frac{1}{2}}^{(1)}\bigl((n+m)f(z)\bigr)
        \end{pmatrix}.
    \end{eqnarray}
    Inserting this into (\ref{theorem:second-column-Y-upper-part:eq2}),
    equation (\ref{theorem:second-column-Y-upper-part:equation})
    is proven in this sector as well.

    Similarly, we can prove
    (\ref{theorem:second-column-Y-upper-part:equation}) in the
    other sectors of the upper part of $U_\delta$.
\end{proof}

And finally, the behavior of the second column of $Y$ in the lower
part of the disk $U_\delta$ is given by the following theorem.

\begin{theorem}\label{theorem:second-column-Y-lower-part}
        Fix $m\in\mathbb Z$. For $z\in U_\delta\cap\mathbb{C}_-$ and $n$
        sufficiently large, the second column of $Y=Y^{(n+m,n)}$ is given by
        \begin{eqnarray}
            \nonumber
            \lefteqn{
            \begin{pmatrix}
                Y_{12}(z) \\[1ex]
                Y_{22}(z)
            \end{pmatrix}
            =
            -z^\alpha e^{-n\frac{V(z)}{2}} \frac{1}{2} \sqrt\pi e^{-\frac{\pi
            i}{4}} e^{(n+m)\frac{\ell}{2}\sigma_3} M(z)} \\
            \label{theorem:second-column-Y-lower-part:equation}
            && \qquad\qquad\qquad\qquad \times\,
            \begin{pmatrix}
                \bigl((n+m)f(z)\bigr)^{1/2} H_{\alpha+\frac{1}{2}}^{(2)}\bigl((n+m)f(z)\bigr) \\[2ex]
                \bigl((n+m)f(z)\bigr)^{1/2} H_{\alpha-\frac{1}{2}}^{(2)}\bigl((n+m)f(z)\bigr)
            \end{pmatrix}.
        \end{eqnarray}
        Here, $H^{(2)}_\nu$ is the Hankel
        function of the second kind of order $\nu$, $f$ is given by {\rm (\ref{definition-f})},
        and $M$ is given by {\rm (\ref{definition-M})}.
\end{theorem}

\begin{proof}
    The proof is similar to the proofs of Theorem \ref{theorem:first-column-Y} and Theorem
    \ref{theorem:second-column-Y-upper-part}.
\end{proof}

\begin{remark}
    By (\ref{RHPYb}), the jump relation for the second column of $Y$
    is
    \begin{equation}\label{jump-relation-second-column}
        \begin{pmatrix}
            Y_{12,+}(x) \\
            Y_{22,+}(x)
        \end{pmatrix} -
        \begin{pmatrix}
            Y_{12,-}(x) \\
            Y_{22,-}(x)
        \end{pmatrix}
        =
        \begin{pmatrix}
            Y_{11}(x) \\
            Y_{21}(x)
        \end{pmatrix} |x|^{2\alpha}e^{-nV(x)},\qquad\mbox{for $x\in\mathbb{R}\setminus\{0\}$.}
    \end{equation}
    For $x\in(0,\delta)$ one can check easily, using (\ref{theorem:first-column-Y}),
    (\ref{theorem:second-column-Y-upper-part:equation}),
    (\ref{theorem:second-column-Y-lower-part:equation}) and \cite[formulas 9.1.3 and
    9.1.4]{AbramowitzStegun} that (\ref{jump-relation-second-column}) is satisfied.
    For $x\in(-\delta,0)$ it follows from (\ref{remark:first-column-Y:equation}),
    (\ref{theorem:second-column-Y-upper-part:equation}), (\ref{theorem:second-column-Y-lower-part:equation})
    and
    \cite[formulas 9.1.3, 9.1.4 and 9.1.39]{AbramowitzStegun} that
    (\ref{jump-relation-second-column}) is satisfied.
\end{remark}

\begin{remark}
    Theorems \ref{theorem:first-column-Y}, \ref{theorem:second-column-Y-upper-part} and
    \ref{theorem:second-column-Y-lower-part} give because of (\ref{RHPYsolution}),
    after straightforward calculations, the
    behavior near the origin
    of the orthogonal polynomials and their Cauchy transforms.
    It has been shown before by Akemann and Fyodorov
    \cite{AkemannFyodorov} that the behavior near the origin of the orthogonal polynomials is given in terms of
    the $J$-Bessel functions $J_{\alpha\pm\frac{1}{2}}$, and that the behavior near the origin
    of their Cauchy transforms is given in terms of the
    Hankel functions $H_{\alpha\pm\frac{1}{2}}^{(1)}$ of the first kind in the upper half-plane,
    and in terms of the Hankel functions $H_{\alpha\pm\frac{1}{2}}^{(2)}$ of the
    second kind in the lower half-plane. However, in \cite{AkemannFyodorov} this was done on a
    physical level of rigor, and under
    the assumption that the eigenvalue density was supported on
    only one interval.
\end{remark}

\section{Proof of Theorem \ref{theorem:WI}--\ref{theorem:WIII}}
    \label{section:proof of theorems}

In this section we prove the universal behavior at the origin of
the spectrum for the three kernels $W_{I,n+m}, W_{II,n+m}$ and
$W_{III,n+m}$, in terms of the Bessel kernels given by Table
\ref{table:limiting-Bessel-kernels}. Similar as in \cite{KV1,KV2},
where we have investigated local eigenvalue correlations, we do
this by using the connection of these kernels with the solution of
the RH problem for $Y$, see (\ref{remark: columnsY:
eq1})--(\ref{remark: columnsY: eq3}), and by using the behavior of
$Y$ near the origin, derived in the previous section.

\medskip

We first need the following lemma's.

\begin{lemma}\label{lemma:estimate-M}
    Let $M$ be the matrix valued function
    given by {\rm (\ref{definition-M})}, and let $\zeta,\eta\in\mathbb{C}$.
    Then, each entry $M_{ij}$ of $M$ satisfies
    \begin{equation}
        M_{ij}\left(\frac{\zeta}{n\psi(0)}\right)-M_{ij}\left(\frac{\eta}{n\psi(0)}\right)
        =O\left(\frac{\zeta-\eta}{n}\right),\qquad \mbox{as $n\to\infty$,}
    \end{equation}
    uniformly for $\zeta$ and $\eta$ in compact subsets of $\mathbb{C}$.
\end{lemma}

\begin{proof}
    Let $\zeta,\eta\in\mathbb{C}$, denote
    $\zeta_n=\frac{\zeta}{n\psi(0)}$ and
    $\eta_n=\frac{\eta}{n\psi(0)}$, and let
    $\gamma$ be a positively oriented simple closed contour in $U_\delta$ going
    around the origin. Then, since $M$ is analytic on $U_\delta$,
    an application of Cauchy's formula
    shows that
    \[
        M_{ij}(\zeta_n)-M_{ij}(\eta_n)=(\zeta_n-\eta_n)\frac{1}{2\pi
        i}\oint_\gamma\frac{M_{ij}(z)}{(z-\zeta_n)(z-\eta_n)}dz,
    \]
    for $\zeta$ and $\eta$ in compact subsets of $\mathbb{C}$ and $n$
    sufficiently large. Since $M_{ij}$ is uniformly bounded in
    $U_\delta$ as $n\to\infty$, see Section \ref{section:behavior of Y near the origin},
    the integral is uniformly bounded
    for $\zeta$ and $\eta$ in compact subsets of $\mathbb{C}$ as $n\to\infty$. This proves the lemma.
\end{proof}

\begin{lemma}\label{lemma:help}
    Fix $m\in\mathbb{Z}$. Let $\zeta,\eta\in\mathbb{C}$, and denote
    \[
        \tilde\zeta_n=(n+m)f\left(\frac{\zeta}{n\psi(0)}\right), \quad\mbox{and}\quad
        \tilde\eta_n=(n+m)f\left(\frac{\eta}{n\psi(0)}\right).
    \]
    Then,
    \begin{equation}\label{lemma:help:equation1}
        \zeta^{-\alpha}\tilde\zeta_n^{1/2}J_{\alpha\pm\frac{1}{2}}(\tilde\zeta_n)=
            \zeta^{-\alpha}(\pi\zeta)^{1/2}J_{\alpha\pm\frac{1}{2}}(\pi\zeta)+O(1/n),
    \end{equation}
    as $n\to\infty$, uniformly for $\zeta$ in compact subsets of
    $\mathbb{C}$. The left hand side of
    {\rm (\ref{lemma:help:equation1})} is uniformly bounded for $\zeta$ in compact subsets of
    $\mathbb{C}$ as $n\to\infty$. Also
    \begin{eqnarray}
        \nonumber
        \lefteqn{
            \zeta^{-\alpha}\tilde\zeta_n^{1/2}J_{\alpha\pm\frac{1}{2}}(\tilde\zeta_n)
                - \eta^{-\alpha}\tilde\eta_n^{1/2}J_{\alpha\pm\frac{1}{2}}(\tilde\eta_n)
        } \\[1ex]
        \label{lemma:help:equation2}
        && \qquad =\,
            \zeta^{-\alpha}(\pi\zeta)^{1/2}J_{\alpha\pm\frac{1}{2}}(\pi\zeta)
            -
            \eta^{-\alpha}(\pi\eta)^{1/2}J_{\alpha\pm\frac{1}{2}}(\pi\eta)
            +O\left(\frac{\zeta-\eta}{n}\right),
    \end{eqnarray}
    as $n\to\infty$, uniformly for $\zeta$ and $\eta$ in compact subsets
    of $\mathbb{C}$.
\end{lemma}

\begin{proof}
    By (\ref{behavior-f}) it follows that
    $\tilde\zeta_n=\pi\zeta\bigl(1+O(1/n)\bigr)$ as
    $n\to\infty$,
    uniformly for $\zeta$ in compact subsets of $\mathbb{C}$.
    Inserting this into the left hand side of (\ref{lemma:help:equation1})
    we easily obtain estimate (\ref{lemma:help:equation1}), cf.\ \cite[Lemma 7.2]{KV2}.
    Since $J_\nu(\zeta)=\zeta^\nu H_\nu(\zeta)$
    with $H_\nu$ entire, see \cite[formula 9.1.10]{AbramowitzStegun}, we have that
    $\zeta^{-\alpha+\frac{1}{2}}J_{\alpha\pm\frac{1}{2}}(\zeta)$ is entire.
    This implies by (\ref{lemma:help:equation1}) that the left hand side of
    (\ref{lemma:help:equation1}) is uniformly bounded for $\zeta$ in compact
    subsets of $\mathbb{C}$ as $n\to\infty$.

    Let $K_1, K_2$ be compact subsets of $\mathbb{C}$, and let
    $\gamma$ be a positively oriented simple closed contour with $K_1$ and
    $K_2$ in its interior. Define
    \begin{equation}\label{lemma:help:eq1}
        q_n(z)=z^{-\alpha}\tilde
        z_n^{1/2}J_{\alpha\pm\frac{1}{2}}(\tilde
        z_n)-z^{-\alpha}(\pi z)^{1/2}J_{\alpha\pm\frac{1}{2}}(\pi
        z),
    \end{equation}
    with $\tilde z_n=(n+m)f(\frac{z}{n\psi(0)})$. Note that $q_n$ is analytic in an open neighborhood
    of the interior of $\gamma$ for $n$ sufficiently large.
    An application of Cauchy's theorem then shows that
    \[
        q_n(\zeta)-q_n(\eta)=(\zeta-\eta)\frac{1}{2\pi
        i}\oint_{\gamma}\frac{q_n(z)}{(z-\zeta)(z-\eta)}dz,
    \]
    for $\zeta\in K_1$ and $\eta\in K_2$, and $n$ sufficiently large.
    Since $q_n(z)=O(1/n)$ as $n\to\infty$ uniformly for $z$ in
    compact subsets of $\mathbb{C}$, see (\ref{lemma:help:equation1}) and
    (\ref{lemma:help:eq1}), and since $\zeta$ and $\eta$ are not close to the contour
    $\gamma$, the lemma is then proven.
\end{proof}

Now, we are ready to prove the universal behavior at the origin of
the spectrum for the kernel $W_{I,n+m}$ .

\begin{varproof}{\bf of Theorem \ref{theorem:WI}.}
    Let $\zeta,\eta\in\mathbb{C}$, denote
    \[
        \zeta_n=\frac{\zeta}{n\psi(0)}, \quad
        \eta_n=\frac{\eta}{n\psi(0)}, \quad
        \tilde\zeta_n=(n+m)f(\zeta_n), \quad \mbox{and} \quad
        \tilde\eta_n=(n+m)f(\eta_n),
    \]
    and let $Y=Y^{(n+m,n)}$. Similar considerations as in
    \cite{KV1,KV2}, using
    (\ref{remark: columnsY: eq1}) and the behavior (\ref{theorem:first-column-Y:equation}) of the first column
    of $Y$ inside the disk $U_\delta$,
    show that,
    \begin{eqnarray}
        \nonumber
        \lefteqn{
            \widehat{W}_{I,n+m}(\zeta,\eta)\equiv
            \gamma_{n+m-1,n}^2\frac{1}{n\psi(0)}W_{I,n+m}(\zeta_n,\eta_n)
        } \\[2ex]
        \nonumber
        &=&
        \frac{1}{-2\pi i(\zeta-\eta)}\det
        \begin{pmatrix}
            Y_{11}(\zeta_n) & Y_{11}(\eta_n) \\
            Y_{21}(\zeta_n) & Y_{21}(\eta_n)
        \end{pmatrix} \\[3ex]
        \nonumber
        &=&
        (n\psi(0))^{2\alpha}e^{\frac{n}{2}(V(\zeta_n)+V(\eta_n))}\frac{1}{2(\zeta-\eta)}
        \\[2ex]
        \label{theorem:WI:eq1a}
        && \quad\times\,
        \det\left[
        M(\zeta_n)
        \begin{pmatrix}
            \zeta^{-\alpha}\tilde\zeta_n^{1/2}J_{\alpha+\frac{1}{2}}(\tilde\zeta_n)
            & 0 \\
            \zeta^{-\alpha}\tilde\zeta_n^{1/2}J_{\alpha-\frac{1}{2}}(\tilde\zeta_n) & 0
        \end{pmatrix}+
        M(\eta_n)
        \begin{pmatrix}
            0 & \eta^{-\alpha}\tilde\eta_n^{1/2}J_{\alpha+\frac{1}{2}}(\tilde\eta_n)
            \\
            0 & \eta^{-\alpha}\tilde\eta_n^{1/2}J_{\alpha-\frac{1}{2}}(\tilde\eta_n)
        \end{pmatrix}\right].
    \end{eqnarray}
    The matrix in the determinant can be written as, cf.\
    \cite{KV1,KV2},
    \begin{eqnarray*}
        \lefteqn{M(\zeta_n)\left[
        \begin{pmatrix}
            \zeta^{-\alpha}\tilde\zeta_n^{1/2}J_{\alpha+\frac{1}{2}}(\tilde\zeta_n)
            & \eta^{-\alpha}\tilde\eta_n^{1/2}J_{\alpha+\frac{1}{2}}(\tilde\eta_n) \\
            \zeta^{-\alpha}\tilde\zeta_n^{1/2}J_{\alpha-\frac{1}{2}}(\tilde\zeta_n)
            & \eta^{-\alpha}\tilde\eta_n^{1/2}J_{\alpha-\frac{1}{2}}(\tilde\eta_n)
        \end{pmatrix}
        \right.} \\[2ex]
        && \left. \qquad\qquad +\, M(\zeta_n)^{-1}(M(\eta_n)-M(\zeta_n))
        \begin{pmatrix}
            0 & \eta^{-\alpha}\tilde\eta_n^{1/2}J_{\alpha+\frac{1}{2}}(\tilde\eta_n)
            \\
            0 & \eta^{-\alpha}\tilde\eta_n^{1/2}J_{\alpha-\frac{1}{2}}(\tilde\eta_n)
        \end{pmatrix}\right].
    \end{eqnarray*}
    Since $\det M\equiv 1$ and each entry of $M$ is uniformly bounded in
    $U_\delta$ as $n\to\infty$, see Section \ref{section:behavior of Y near the origin},
    each entry of $M(\zeta_n)^{-1}$ is
    uniformly bounded for $\zeta$ in compact subsets of $\mathbb{C}$
    as $n\to\infty$. Using Lemma \ref{lemma:estimate-M} and the fact that
    $\eta^{-\alpha}\tilde\eta_n^{1/2}J_{\alpha\pm\frac{1}{2}}(\tilde\eta_n)$
    is uniformly bounded for $\eta$ in compact subsets of $\mathbb{C}$ as $n\to\infty$, see Lemma
    \ref{lemma:help}, we then obtain
    \[
        M(\zeta_n)^{-1}(M(\eta_n)-M(\zeta_n))
        \begin{pmatrix}
            0 & \eta^{-\alpha}\tilde\eta_n^{1/2}J_{\alpha+\frac{1}{2}}(\tilde\eta_n)
            \\
            0 & \eta^{-\alpha}\tilde\eta_n^{1/2}J_{\alpha-\frac{1}{2}}(\tilde\eta_n)
        \end{pmatrix}=
        \begin{pmatrix}
            0 & O\left(\frac{\zeta-\eta}{n}\right) \\[1ex]
            0 & O\left(\frac{\zeta-\eta}{n}\right)
        \end{pmatrix}.
    \]
    Using the facts that $\det M\equiv 1$ and that
    $\zeta^{-\alpha}\tilde\zeta_n^{1/2}J_{\alpha\pm\frac{1}{2}}(\tilde\zeta_n)$
    is uniformly bounded for $\zeta$ in compact subsets of $\mathbb{C}$ as $n\to\infty$,
    see Lemma \ref{lemma:help}, we then find
    \begin{eqnarray}
        \nonumber
        \lefteqn{
            \widehat{W}_{I,n+m}(\zeta,\eta)=(n\psi(0))^{2\alpha}e^{\frac{n}{2}(V(\zeta_n)+V(\eta_n))}
        } \\[2ex]
        \label{theorem:WI:eq1}
        && \times\left[
        \frac{1}{2(\zeta-\eta)}
        \det
        \begin{pmatrix}
            \zeta^{-\alpha}\tilde\zeta_n^{1/2}J_{\alpha+\frac{1}{2}}(\tilde\zeta_n)
            & \eta^{-\alpha}\tilde\eta_n^{1/2}J_{\alpha+\frac{1}{2}}(\tilde\eta_n) \\
            \zeta^{-\alpha}\tilde\zeta_n^{1/2}J_{\alpha-\frac{1}{2}}(\tilde\zeta_n)
            & \eta^{-\alpha}\tilde\eta_n^{1/2}J_{\alpha-\frac{1}{2}}(\tilde\eta_n)
        \end{pmatrix}+O(1/n)
        \right].
    \end{eqnarray}
    We can now replace $z^{-\alpha}\tilde z_n^{1/2}J_{\alpha\pm\frac{1}{2}}(\tilde z_n)$
    by $z^{-\alpha}(\pi z)^{1/2}J_{\alpha\pm\frac{1}{2}}(\pi z)$ for
    $z=\zeta,\eta$, and obtain the limiting Bessel kernel
    $\mathbb{J}_{\alpha,I}(\zeta,\eta)$ given in Table \ref{table:limiting-Bessel-kernels}.
    However, then we make an error which does not
    hold uniformly for $\zeta$ and $\eta$ close to each other. To
    solve this problem we will work as in \cite{KV1,KV2}. We subtract the second column
    in the determinant from the first one. From (\ref{lemma:help:equation2}) and the fact
    that $\eta^{-\alpha}\tilde\eta_n^{1/2}J_{\alpha\pm\frac{1}{2}}(\tilde\eta_n)$ is uniformly
    bounded for $\eta$ in compact subsets of $\mathbb{C}$ as $n\to\infty$, the term inside the
    brackets in (\ref{theorem:WI:eq1}) is then given by
    \[
        \frac{1}{2(\zeta-\eta)}
        \det
        \begin{pmatrix}
            \zeta^{-\alpha}(\pi\zeta)^{1/2}J_{\alpha+\frac{1}{2}}(\pi\zeta)-
                \eta^{-\alpha}(\pi\eta)^{1/2}J_{\alpha+\frac{1}{2}}(\pi\eta)
            & \eta^{-\alpha}\tilde\eta_n^{1/2}J_{\alpha+\frac{1}{2}}(\tilde\eta_n) \\
            \zeta^{-\alpha}(\pi\zeta)^{1/2}J_{\alpha-\frac{1}{2}}(\pi\zeta)-
                \eta^{-\alpha}(\pi\eta)^{1/2}J_{\alpha-\frac{1}{2}}(\pi\eta)
            & \eta^{-\alpha}\tilde\eta_n^{1/2}J_{\alpha-\frac{1}{2}}(\tilde\eta_n)
        \end{pmatrix}+O(1/n).
    \]
    Using the fact that
    \[
        \eta^{-\alpha}\tilde\eta_n^{1/2}J_{\alpha\pm\frac{1}{2}}(\tilde\eta_n)=
            \eta^{-\alpha}(\pi\eta)^{1/2}J_{\alpha\pm\frac{1}{2}}(\pi\eta)+O(1/n),
    \]
    and the fact that
    \[
        \frac{\zeta^{-\alpha}(\pi\zeta)^{1/2}J_{\alpha\pm\frac{1}{2}}(\pi\zeta)-
            \eta^{-\alpha}(\pi\eta)^{1/2}J_{\alpha\pm\frac{1}{2}}(\pi\eta)}{\zeta-\eta}
    \]
    remains bounded for $\zeta$ and $\eta$ in compact subsets of
    $\mathbb{C}$, which follows since $z^{-\alpha}(\pi z)^{1/2}J_{\alpha\pm\frac{1}{2}}(z)$
    is entire, we then easily obtain that the term inside the
    brackets in (\ref{theorem:WI:eq1}) is given by
    \begin{equation}\label{theorem:WI:eq2a}
        \frac{1}{2(\zeta-\eta)}
        \det
        \begin{pmatrix}
            \zeta^{-\alpha}(\pi\zeta)^{1/2}J_{\alpha+\frac{1}{2}}(\pi\zeta)
            & \eta^{-\alpha}(\pi\eta)^{1/2}J_{\alpha+\frac{1}{2}}(\pi\eta) \\
            \zeta^{-\alpha}(\pi\zeta)^{1/2}J_{\alpha-\frac{1}{2}}(\pi\zeta)
            & \eta^{-\alpha}(\pi\eta)^{1/2}J_{\alpha-\frac{1}{2}}(\pi\eta)
        \end{pmatrix}+O(1/n).
    \end{equation}
    The first term in (\ref{theorem:WI:eq2a}) is exactly
    the limiting Bessel kernel $\mathbb{J}_{\alpha,I}(\zeta,\eta)$, see
    Table \ref{table:limiting-Bessel-kernels}. From (\ref{theorem:WI:eq1}) and
    (\ref{theorem:WI:eq2a}) we then
    obtain
    \begin{equation}\label{theorem:WI:eq2}
        \widehat{W}_{I,n+m}(\zeta,\eta)=(n\psi(0))^{2\alpha}e^{\frac{n}{2}(V(\zeta_n)+V(\eta_n))}
        \Bigl(\mathbb{J}_{\alpha,I}(\zeta,\eta)+O(1/n)\Bigr),
    \end{equation}
    as $n\to\infty$, uniformly for $\zeta$ and $\eta$ in bounded
    subsets of $\mathbb{C}$. Note that
    \begin{eqnarray*}
        e^{\frac{n}{2}(V(\zeta_n)+V(\eta_n))} &=&
            e^{nV(0)+\frac{V'(0)}{2\psi(0)}(\zeta+\eta)+O(1/n)}
            \\[1ex]
        &=&
        e^{nV(0)}e^{\frac{V'(0)}{2\psi(0)}(\zeta+\eta)}(1+O(1/n)),\qquad
        \mbox{as $n\to\infty$,}
    \end{eqnarray*}
    uniformly for $\zeta$ and $\eta$ in compact subsets of $\mathbb{C}$.
    Inserting this into (\ref{theorem:WI:eq2}), and
    using the fact that $\mathbb{J}_{\alpha,I}(\zeta,\eta)$ is
    bounded for $\zeta$ and
    $\eta$ in compact subsets of $\mathbb{C}$
    the theorem is then proven.
\end{varproof}

In order to prove Theorem \ref{theorem:WII} we also need the
following lemma, which is analogous to Lemma \ref{lemma:help}.

\begin{lemma}\label{lemma:help2}
    Fix $m\in\mathbb{Z}$. Let $\zeta\in\mathbb{C}_+$, and denote
    $\tilde\zeta_n=(n+m)f(\frac{\zeta}{n\psi(0)})$.
    Then,
    \begin{equation}\label{lemma:help2:equation}
        \zeta^{\alpha}\tilde\zeta_n^{1/2}H^{(1)}_{\alpha\pm\frac{1}{2}}(\tilde\zeta_n)=
            \zeta^{\alpha}(\pi\zeta)^{1/2}H^{(1)}_{\alpha\pm\frac{1}{2}}(\pi\zeta)+O(1/n),
    \end{equation}
    as $n\to\infty$, uniformly for $\zeta$ in compact subsets of
    $\mathbb{C}_+$. The left hand side of
    {\rm (\ref{lemma:help2:equation})} is uniformly bounded for $\zeta$ in compact subsets of
    $\mathbb{C}_+$ as $n\to\infty$.
\end{lemma}

\begin{proof}
    Recall, cf.\ the proof of Lemma \ref{lemma:help}, that $\tilde\zeta_n=(\pi\zeta)(1+O(1/n))$.
    Inserting this into the left hand side of
    (\ref{lemma:help2:equation}) and using the fact that
    $H_{\alpha\pm\frac{1}{2}}^{(1)}$ is analytic on
    $\mathbb{C}\setminus(-\infty,0]$ we easily obtain estimate
    (\ref{lemma:help2:equation}). Since
    $H^{(1)}_{\alpha\pm\frac{1}{2}}$ is analytic on
    $\mathbb{C}\setminus(-\infty,0]$, we have that
    $\zeta^{\alpha}(\pi\zeta)^{1/2}H^{(1)}_{\alpha\pm\frac{1}{2}}(\pi\zeta)$
    is bounded for $\zeta$ in compact subsets of
    $\mathbb{C}\setminus(-\infty,0]$, and thus in particular in compact subsets of
    $\mathbb{C}_+$. Together with estimate
    (\ref{lemma:help2:equation}) this implies that the left hand
    side of (\ref{lemma:help2:equation}) remains uniformly bounded
    for $\zeta$ in compact subsets of $\mathbb{C}_+$ as $n\to\infty$.
\end{proof}

\begin{varproof}{\bf of Theorem \ref{theorem:WII}.}
    Let $\zeta\in\mathbb{C}_+,\, \eta\in\mathbb{C}$, denote
    \[
        \zeta_n=\frac{\zeta}{n\psi(0)},\quad
        \eta_n=\frac{\eta}{n\psi(0)},\quad
        \tilde\zeta_n=(n+m)f(\zeta_n),\quad \mbox{and}\quad
        \tilde\eta_n=(n+m)f(\eta_n),
    \]
    and let $Y=Y^{(n+m,n)}$.
    Instead of equation (\ref{theorem:WI:eq1a}) we obtain from (\ref{remark: columnsY:
    eq2}), from the behavior
    (\ref{theorem:first-column-Y:equation}) of the first column of $Y$ inside
    $U_\delta$,
    and from the behavior
    (\ref{theorem:second-column-Y-upper-part:equation}) of the second column of $Y$ in the upper part
    of $U_\delta$,
    \begin{eqnarray}
        \nonumber
        \lefteqn{
            \widehat W_{II,n+m}(\zeta,\eta)\equiv \gamma_{n+m-1,n}^2
            \frac{\zeta-\eta}{n\psi(0)}W_{II,n+m}(\zeta_n,\eta_n)
            }
            \\[2ex]
        \nonumber
        &=&
            \frac{1}{-2\pi i}\det
            \begin{pmatrix}
                Y_{12}(\zeta_n) & Y_{11}(\eta_n) \\
                Y_{22}(\zeta_n) & Y_{21}(\eta_n)
            \end{pmatrix} \\[3ex]
        \nonumber
        &=&
            \frac{1}{4} e^{-\frac{n}{2}(V(\zeta_n)-V(\eta_n))}  \\[2ex]
        \nonumber
        &&
            \quad\times\,
            \det\left[M(\zeta_n)
            \begin{pmatrix}
                \zeta^{\alpha}\tilde\zeta_n^{1/2}H_{\alpha+\frac{1}{2}}^{(1)}(\tilde\zeta_n)
                & 0 \\[2ex]
                \zeta^{\alpha}\tilde\zeta_n^{1/2}H_{\alpha-\frac{1}{2}}^{(1)}(\tilde\zeta_n)
                & 0
            \end{pmatrix}+M(\eta_n)
            \begin{pmatrix}
                0 & \eta^{-\alpha}\tilde\eta_n^{1/2}J_{\alpha+\frac{1}{2}}(\tilde\eta_n) \\[2ex]
                0 & \eta^{-\alpha}\tilde\eta_n^{1/2}J_{\alpha-\frac{1}{2}}(\tilde\eta_n)
            \end{pmatrix}\right].
    \end{eqnarray}
    We now rewrite the matrix in the determinant as was done in the proof of Theorem
    \ref{theorem:WI}. Using also the fact that
    $\zeta^{\alpha}\tilde\zeta_n^{1/2}H_{\alpha\pm\frac{1}{2}}^{(1)}(\tilde\zeta_n)$
    is uniformly bounded for $\zeta$ in
    compact subsets of $\mathbb{C}_+$ as $n\to\infty$, see Lemma \ref{lemma:help2},
    we obtain instead of equation (\ref{theorem:WI:eq1}), in a similar fashion, the following
    \begin{eqnarray}
        \nonumber
        \lefteqn{
            \widehat W_{II,n+m}(\zeta,\eta) =
                e^{-\frac{n}{2}(V(\zeta_n)-V(\eta_n))}
        } \\[2ex]
        \label{theorem:WII:eq1}
        && \qquad\qquad\times\,
            \left[\frac{1}{4} \det
            \begin{pmatrix}
                \zeta^{\alpha}\tilde\zeta_n^{1/2}H_{\alpha+\frac{1}{2}}^{(1)}(\tilde\zeta_n)
                & \eta^{-\alpha}\tilde\eta_n^{1/2}J_{\alpha+\frac{1}{2}}(\tilde\eta_n)\\[2ex]
                \zeta^{\alpha}\tilde\zeta_n^{1/2}H_{\alpha-\frac{1}{2}}^{(1)}(\tilde\zeta_n)
                & \eta^{-\alpha}\tilde\eta_n^{1/2}J_{\alpha-\frac{1}{2}}(\tilde\eta_n)
            \end{pmatrix}+O(1/n)\right],
    \end{eqnarray}
    as $n\to\infty$, uniformly for $\zeta$ and $\eta$ in compact subsets of
    $\mathbb{C}_+$ and $\mathbb{C}$, respectively.
    We now insert the fact that, see Lemma \ref{lemma:help2}
    \[
        \zeta^{\alpha}\tilde\zeta_n^{1/2}H_{\alpha\pm\frac{1}{2}}^{(1)}(\tilde\zeta_n)=
            \zeta^\alpha
            (\pi\zeta)^{1/2}H_{\alpha\pm\frac{1}{2}}^{(1)}(\pi\zeta)+O(1/n),
    \]
    into (\ref{theorem:WII:eq1}), and use the fact that $\eta^{-\alpha}\tilde\eta_n^{1/2}
    J_{\alpha\pm\frac{1}{2}}(\tilde\eta_n)$ is uniformly bounded for $\eta$ in compact
    subsets of $\mathbb{C}$ as $n\to\infty$, see
    Lemma \ref{lemma:help}, to obtain
    \begin{eqnarray}
        \nonumber
        \lefteqn{
            \widehat W_{II,n+m}(\zeta,\eta) =
                e^{-\frac{n}{2}(V(\zeta_n)-V(\eta_n))}
        } \\[2ex]
        \label{theorem:WII:eq2}
        && \qquad\qquad\times\,
            \left[\frac{1}{4}\det
            \begin{pmatrix}
                \zeta^{\alpha}(\pi\zeta)^{1/2}H_{\alpha+\frac{1}{2}}^{(1)}(\pi\zeta)
                & \eta^{-\alpha}\tilde\eta_n^{1/2}J_{\alpha+\frac{1}{2}}(\tilde\eta_n)\\[2ex]
                \zeta^{\alpha}(\pi\zeta)^{1/2}H_{\alpha-\frac{1}{2}}^{(1)}(\pi\zeta)
                & \eta^{-\alpha}\tilde\eta_n^{1/2}J_{\alpha-\frac{1}{2}}(\tilde\eta_n)
            \end{pmatrix}+O(1/n)\right].
    \end{eqnarray}
    Inserting, see Lemma \ref{lemma:help}
    \[
        \eta^{-\alpha}\tilde\eta_n^{1/2}
        J_{\alpha\pm\frac{1}{2}}(\tilde\eta_n)=\eta^{-\alpha}(\pi\eta)^{1/2}J_{\alpha\pm\frac{1}{2}}(\pi\eta)+O(1/n),
    \]
    into (\ref{theorem:WII:eq2}), and using the fact that
    $\zeta^{\alpha}(\pi\zeta)^{1/2}H_{\alpha\pm\frac{1}{2}}^{(1)}(\pi\zeta)$ is uniformly bounded
    for $\zeta$ in compact subsets of $\mathbb{C}_+$ as $n\to\infty$, we then
    obtain
    \begin{eqnarray}
        \nonumber
        \lefteqn{
            \widehat W_{II,n+m}(\zeta,\eta)
        } \\[1ex]
        \nonumber
        &=& e^{-\frac{n}{2}(V(\zeta_n)-V(\eta_n))}
            \left[\frac{1}{4}\det
            \begin{pmatrix}
                \zeta^{\alpha}(\pi\zeta)^{1/2}H_{\alpha+\frac{1}{2}}^{(1)}(\pi\zeta)
                & \eta^{-\alpha}(\pi\eta)^{1/2}J_{\alpha+\frac{1}{2}}(\pi\eta) \\[2ex]
                \zeta^{\alpha}(\pi\zeta)^{1/2}H_{\alpha-\frac{1}{2}}^{(1)}(\pi\zeta)
                & \eta^{-\alpha}(\pi\eta)^{1/2}J_{\alpha-\frac{1}{2}}(\pi\eta)
            \end{pmatrix}+O(1/n)\right] \\[3ex]
        \label{theorem:WII:eq3}
        &=&
        e^{-\frac{n}{2}(V(\zeta_n)-V(\eta_n))}\left(
        (\zeta-\eta)\mathbb{J}_{\alpha,II}^+(\zeta,\eta)+O(1/n)\right),
    \end{eqnarray}
    as $n\to\infty$, uniformly for $\zeta$ and $\eta$ in compact
    subsets of $\mathbb{C}_+$ and $\mathbb{C}$, respectively.
    In (\ref{theorem:WII:eq3}), $\mathbb{J}_{\alpha,II}^+(\zeta,\eta)$ is the Bessel kernel given in
    Table \ref{table:limiting-Bessel-kernels}. Note that
    \begin{eqnarray*}
        e^{-\frac{n}{2}\left(V(\zeta_n)-V(\eta_n)\right)}
        &=&
            e^{-\frac{V'(0)}{2\psi(0)}(\zeta-\eta)+O(1/n)}, \\[1ex]
        &=&
            e^{-\frac{V'(0)}{2\psi(0)}(\zeta-\eta)}(1+O(1/n)),\qquad
            \mbox{as $n\to\infty$,}
    \end{eqnarray*}
    uniformly for $\zeta$ and $\eta$ in compact subsets of $\mathbb{C}_+$ and
    $\mathbb{C}$, respectively. Inserting this into (\ref{theorem:WII:eq3})
    and using the fact that $(\zeta-\eta)\mathbb{J}_{\alpha,II}^+(\zeta,\eta)$ is bounded for
    $\zeta$ and $\eta$ in compact subsets of $\mathbb{C}_+$ and
    $\mathbb{C}$, respectively,
    the first part of the theorem is then proven.

    The second part
    of the theorem can be treated in the same way, using the behavior
    (\ref{theorem:second-column-Y-lower-part:equation}) of the second column of
    $Y$ in the lower part of the disk $U_\delta$,
    instead of in the upper part of the disk.
\end{varproof}

We leave it as an exercise for the careful reader to prove the
universal behavior at the origin of the spectrum for the kernel
$W_{III,n+m}$.

\begin{varproof}{\bf of Theorem \ref{theorem:WIII}.}
    The proof is similar to the proofs of Theorem \ref{theorem:WI} and
    Theorem \ref{theorem:WII}.
\end{varproof}

\subsection*{Acknowledgements}

I thank Arno Kuijlaars for careful reading, as well as for useful
discussions and comments. I am also grateful to Yan Fyodorov and
Eugene Strahov for sending me the recent version of their
manuscript ``\,Universal results for correlations of
characteristic polynomials: Riemann-Hilbert approach".

\end{document}